%% file: iclr2025_conference.tex
\documentclass{article} 
\usepackage{iclr2025_conference,times}

\input{math_commands.tex}

\usepackage{hyperref}
\usepackage{url}
\usepackage{booktabs}       
\usepackage{amsfonts}       
\usepackage{nicefrac}       
\usepackage{microtype}      
\usepackage{graphicx}
\usepackage{wrapfig}
\usepackage{lipsum}
\usepackage{enumitem}
\usepackage{booktabs}
\usepackage{amsmath, amssymb}
\usepackage{multirow}
\usepackage{multirow}

\newcommand{\modelEmoji}{\includegraphics[height=1.1em,trim=0 .9em 0 0]{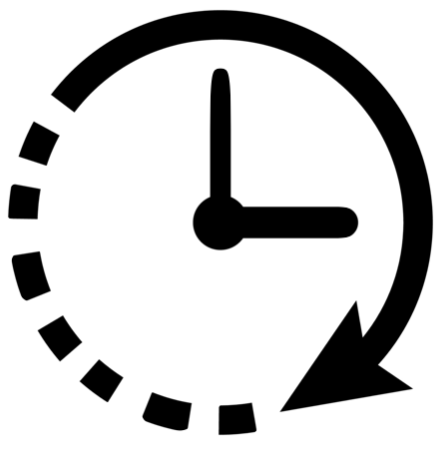}}

\title{\modelEmoji \ VersiCode: Towards Version-controllable Code Generation}

\newcommand{\monash}{\spadesuit}
\newcommand{\njupt}{\heartsuit}
\newcommand{\tiktok}{\clubsuit}
\newcommand{\data}{\diamondsuit}

\author{Tongtong Wu$^{\monash}$\ \ \ \ \ Weigang Wu$^{\njupt}$\ \ \ \ \  Xingyu Wang$^{\njupt}$\ \ \ \ \  Kang Xu$^\njupt$\ \ \ \ \ Suyu Ma$^\data$\ \ \ \ \ Bo Jiang$^\tiktok$ \\
\textbf{Ping Yang}$^{\tiktok}$\ \ \ \ \ \textbf{Zhenchang Xing}$^{\data}$\ \ \ \ \ \textbf{Yuan-Fang Li}$^{\monash}$\ \ \ \ \ \textbf{Gholamreza Haffari}$^{\monash}$ \\ [4pt]
$^{\monash}$Monash University, Australia; $^{\njupt}$Nanjing University of Posts and Telecommunications, China;\\[4pt]
$^{\tiktok}$ByteDance Ltd., China; $^{\data}$CSIRO’s Data61, Australia;  \\[4pt]
$^{\monash}$\texttt{\{first-name.last-name\}@monash.edu}\\[4pt]
}

\iclrfinalcopy 
\begin{document}

\maketitle

\begin{abstract}
Large Language Models (LLMs) have made tremendous strides in code generation, but existing research fails to account for the dynamic nature of software development, marked by frequent library updates. 
This gap significantly limits LLMs' deployment in realistic settings. 
In this paper, we propose two novel tasks aimed at bridging this gap: version-specific code completion (VSCC) and version-aware code migration (VACM). 
In conjunction, we introduce VersiCode, a comprehensive Python dataset specifically designed to evaluate LLMs on these two tasks, together with a novel evaluation metric, Critical Diff Check (CDC@1), which assesses code generation against evolving API requirements. 
We conduct an extensive evaluation on VersiCode, which reveals that version-controllable code generation is indeed a significant challenge, even for GPT-4o and other strong frontier models. 
We believe the novel tasks, dataset and metric open up a new, important research direction that will further enhance LLMs' real-world applicability. 
The code and resources can be found at
\url{https://wutong8023.site/VersiCode/}. 

\end{abstract}

\input{sections/1_intro}

\input{sections/2_task_definition}

\input{sections/3_1_token_level_completion}

\input{sections/3_2_block_level_completion}

\input{sections/3_3_migration}

\input{sections/4_discussion}

\input{sections/5_related_work}

\input{sections/6_conclusion}

\bibliography{iclr2025_conference}
\bibliographystyle{iclr2025_conference}


\appendix
\renewcommand{\thesection}{\Alph{section}}
\newpage

\input{sections/A_dataset_construction}

\input{sections/B_1_dataset_statistics}

\input{sections/B_2_related_dataset}

\input{sections/C_additional_experiments}

\input{sections/D_metric_design}

\input{sections/E_instance_example}

\input{sections/F_prompt_template}

\input{sections/G_error_analysis}

\end{document}

%% file: math_commands.tex
\usepackage{amsmath,amsfonts,bm}

\def\eqref#1{equation~\ref{#1}}

\def\1{\bm{1}}

\DeclareMathAlphabet{\mathsfit}{\encodingdefault}{\sfdefault}{m}{sl}
\SetMathAlphabet{\mathsfit}{bold}{\encodingdefault}{\sfdefault}{bx}{n}



%% file: sections/1_intro.tex
\section{Introduction}
\label{sect:intro}
Large Language Models (LLMs), including OpenAI's GPT series~\citep{abs-2303-08774,gpt35,gpt4o} and specialized variants such as CodeLLaMA~\citep{abs-2308-12950}, have demonstrated significant advancements in code generation tasks. Typically evaluated using benchmarks such as HumanEval \citep{abs-2107-03374} and MBPP \citep{abs-2108-07732}, these models are measured on tasks that assume code generation is a \emph{static} activity. However, the reality of software development is inherently dynamic, characterized by frequent updates to software libraries, which necessitate adjustments to API interfaces. This evolving landscape raises crucial challenges for LLMs, particularly their ability to generate code that is functional for different, specific library versions. This dynamic nature of software development leads us to ask the following questions:
\begin{itemize}[noitemsep,nolistsep]
    \item How reliably can LLMs generate code compatible with specific library versions?
    \item How effectively can LLMs adapt code for API changes across library versions?
\end{itemize}

Existing benchmarks~\citep{jiang2024survey,abs-2403-14734,abs-2403-00338}, which are oblivious to version-specific dynamics, do not address these challenges. They fall short of simulating the continuous version management activities undertaken by developers who ensure the software remains functional across updates. The static nature of existing benchmarks represents a significant barrier to the practical deployment of LLMs in professional environments, where handling version-specific dependencies is critical~\citep{ZhangZWTLX20,ZhangY00RG21,DilharaKD21,Liu0YFG21,Wang0LC20,VadlamaniKC21,HaryonoT0LJ21}.

\begin{figure*}
    \centering
    \includegraphics[width=.98\linewidth, bb=0 0 577.5 225.5]{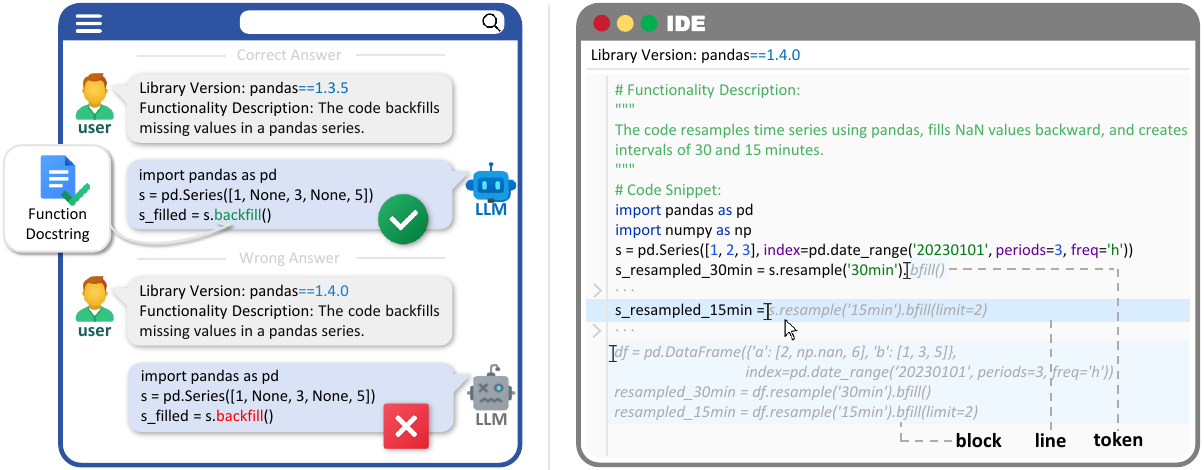}
    \caption{{\small Two motivating scenarios for version-controllable code generation: (left) Interacting with LLMs in a browser, where slight query changes lead to incorrect answers, and (right) Programming in an IDE, explicitly specifying the version of dependency libraries.\label{fig:example}}}
\end{figure*}

To bridge this gap, we propose two novel tasks aimed at evaluating LLMs' version-controllable code generation capabilities, namely version-specific code completion (VSCC) and version-aware code migration (VACM). 
These tasks are crafted to mimic real-world software development scenarios, motivated in Figure~\ref{fig:example}, requiring models to generate code that not only is syntactically correct but also adheres to version-specific API contracts~\citep{ZhangZWTLX20,ZhangY00RG21,DilharaKD21,Liu0YFG21,Wang0LC20,VadlamaniKC21,HaryonoT0LJ21}.
Moreover, we introduce VersiCode, the first dataset specifically designed for these two tasks. VersiCode includes data spanning over 300 Python libraries and more than 2,000 versions across 9 years. It has undergone a careful curation process to ensure high quality. Thus, VersiCode provides a comprehensive and robust testbed for assessing LLMs under realistic conditions. 
Furthermore, we propose a new evaluation metric, CDC (Critical Diff Check), which enhances traditional code similarity metrics by incorporating considerations for API usage, parameter handling, and deprecated features management. This metric offers a more granular assessment of a model’s ability to navigate the complexities of evolving software libraries.

Our extensive testing of strong frontier models like GPT-4o and LLaMA3~\citep{llama3modelcard} reveals significant challenges in version-aware code generation tasks. We uncover that (1) LLMs often retain outdated programming knowledge, particularly concerning version-specific information. (2) Conventional metrics used for evaluating code generation do not effectively capture the nuances of version sensitivity. (3) While leveraging context from various library versions can be beneficial, its utility can be limited.
Guided by these insights, we suggest strategies, such as targeted pretraining, continual learning, and refined evaluation methods, for improving LLMs' version-controlled code generation capabilities.

Our contributions are summarized as follows:
\begin{itemize}[noitemsep,nolistsep]
\item We propose two novel and important yet under-explored tasks in code generation, namely version-specific code completion and version-aware code migration.
\item We introduce VersiCode, a comprehensive, well-documented and \emph{versioned} dataset, accompanied by a subset annotated with executable test cases.
\item We introduce Critical Diff Check, a new metric that extends traditional code similarity metrics by checking syntactic validity, API usage, parameter matching, the use of `with` statements, and correct keyword arguments in the generated code, providing a more detailed evaluation of version-specific code generation.
\item Our thorough experiments provide valuable insights and directions for future research in this critical area of software development.

\end{itemize}

%% file: sections/2_task_definition.tex
\section{Version-controllable Code Generation}
\label{sect:task_definition}
VersiCode is a large-scale code generation benchmark dataset focusing on evolving library dependen-
cies. 
We curated our dataset by initially selecting popular Python repositories from GitHub, confirmed by their star ratings, and ensured they were permissively licensed. For each library, we compiled data from three main sources: (1) Library Source Code, extracting all pip-installable versions and official API usage examples from docstrings; (2) Downstream Application Code, sourcing from top-tier research papers spanning ten years to capture evolving libraries; (3) Stack Overflow, retrieving FAQs that mention specific library versions. 
We present the dataset statistics, construction process and examples in detail in Appendix~\ref{subsect:back_compatib}.

\begin{figure*}[t]
    \centering
    \includegraphics[width=1.0\linewidth, bb=0 0 780 246]{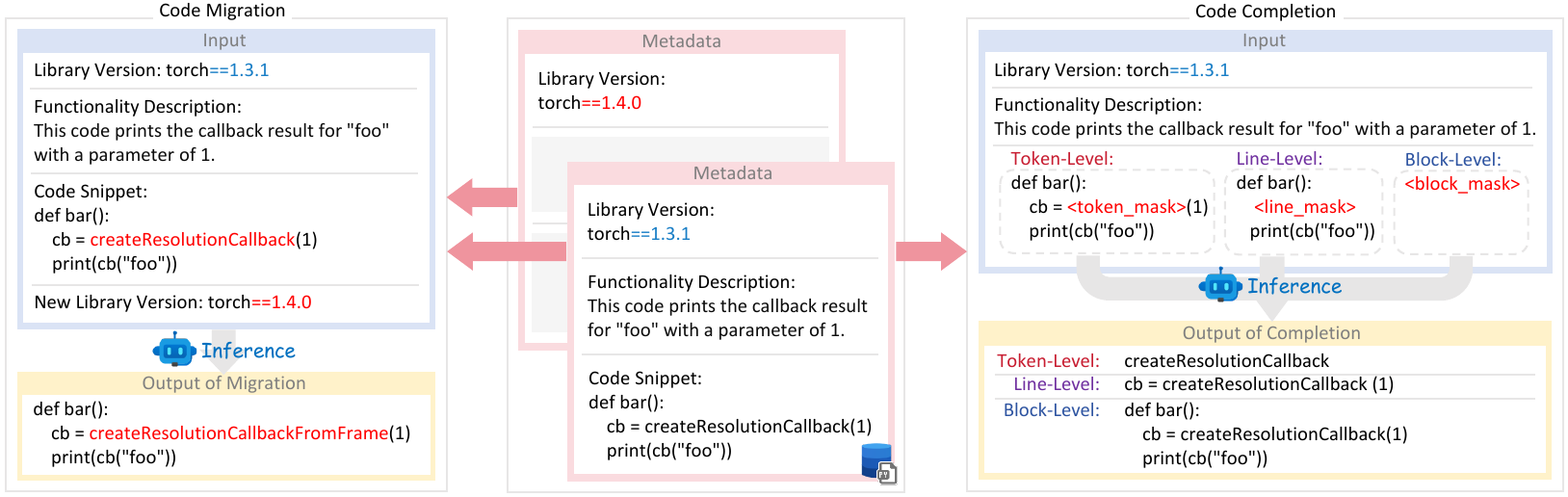}
    \caption{{\small The post-processing pipeline transforms metadata into specific tasks and the running example per task: (left) Leveraging pairs of metadata that share the same functionality but different library versions to construct block-level code migration instances; (right) Utilizing each metadata sample, masking version-sensitive content to create multi-granularity code completion instances.\label{fig:postprocess}}}
    
\end{figure*}

\label{subsect:back_compatib}
As shown in Figure~\ref{fig:postprocess}, we define a \emph{meta-instance} as $m = [l; v; d; c] \in \mathcal{M} $, where $ l $, $ v $, $ d $, and $ c $ represent the library name, version, functionality description, and code snippet, respectively. 
Consider an API $a$ added to library $l$ in version $v_s$ and deprecated in version $v_e$, and is active in the intermediate version $v_m$ where $s \leq m \leq e$. We refer to the interval $[s,e)$ as the \emph{lifecycle} of $a$. To analyze model performance in detail, we assess how up-to-date each LLM is concerning newly added or deprecated APIs per version. We compare the source code between any two consecutive versions of each library to detect changes in API or method names. Based on the detection results, we label the source code as follows: ``addition'' indicates an API newly added in the current version and still applicable in subsequent versions; ``deprecation'' indicates the current version is the last usable version for the API; and ``general'' indicates the API usage method is inherited from the previous version.


We introduce the two novel version-controllable code generation tasks below. 

\noindent\textbf{Version-Specific Code Completion (VSCC)}: Given a meta-instance $ m_i $, the input is $ x = [l_i; v_i; d_i; c'_i] $, where $ c'_i $ is the code snippet $ c_i $ with selective masking, replacing the library- and version-sensitive contents with a special token. Depending on the length of the masked contents, the special token is defined as ``[token-mask]'', ``[line-mask]'', or ``[block-mask]'', reflecting code completion on different granularity levels. The output $ y $ is the masked content, typically containing function names or variables.

\noindent\textbf{Version-Aware Code Migration (VACM)}: Given a pair of meta-instances $ (m_i, m_j | l_i == l_j, d_i == d_j, v_i \neq v_j) $, the input $ x = [l_i; v_i; d_i; c_i; v_j] $, and the output $ y = c_j $. Note that version editing may require refactoring of the code structure, making it difficult to format as detailed as in token-level or line-level completion. Additionally, depending on the numerical relationship between $ v_i $ and $ v_j $, various scenarios arise, such as editing from an old version to a new version, or vice versa. Data statistics are detailed in Appendix~\ref{apd:data_statistics}


%% file: sections/3_1_token_level_completion.tex
\section{Token-level Version-specific Code Completion}
\label{sect:token_level_completion}
In code generation that targets a specific version of a third-party library, the version-related changes usually involve updates to identifiers, such as the addition, removal, or renaming of classes, functions, and parameters. 
The token-level code completion task for a specified version, predicting the evolving identifiers identified in real code, is a fundamental and direct way to evaluate LLMs to generate code for specific versions.
We begin our research by addressing the following three research questions: 
(1) How well do LLMs perform on code completion tasks that involve version-specific library usage compared to other benchmarks like HumanEval and MBPP?
(2) How do LLMs handle new, deprecated, and intermediate versions of libraries in code completion tasks?
(3) How does the performance of LLMs in code completion change over time with different library versions?

\subsection{Experiment Setup}
\label{subsect:setup}

\noindent\textbf{Models}: 
We benchmarked VersiCode against popular open-domain LLMs and dedicated code-LLMs, including variant families such as GPT~\citep{abs-2303-08774,gpt35,gpt4o}, LLaMa~\citep{abs-2307-09288}, Mistral~\citep{abs-2310-06825}, CodeLLaMa~\citep{abs-2308-12950}, CodeQwen~\citep{qwen}, CodeGemma~\citep{codegemma_2024}, StarCoder~\citep{abs-2402-19173}, Deepseek-Coder~\citep{abs-2401-14196}, and WizardCoder~\citep{abs-2306-08568}. For smaller open-source models (e.g., \textless 20B parameters), we downloaded them from HuggingFace~\footnote{\url{https://huggingface.co/models}} and deployed them locally for inference. For larger models, such as LLaMa3 70B~\citep{llama3modelcard} and GPT-4o~\citep{gpt4o}, we used their online APIs~\footnote{\url{https://together.ai}}~\footnote{\url{https://openai.com/}} for inference.

\noindent\textbf{Data Preparation}: 
 Each instance in VersiCode is tagged with its data source (library source code, downstream applications, or Stack Overflow), feature type (addition, deprecation, or general), and release time, allowing for more detailed performance analysis. We randomly selected 2,000 instances for token-level code completion. (see Appendix~\ref{subapd:data_prepare}).

\noindent\textbf{Baseline Dataset}: To assess the difficulty of VersiCode, we compared it with two well-known code generation datasets, HumanEval~\citep{LiuXW023} and MBPP~\citep{jiang2024survey}, and observed the overall performance of models. HumanEval~\citep{LiuXW023} measures functional correctness in synthesizing programs from docstrings with 164 original problems, resembling simple software interview questions. MBPP~\citep{abs-2108-07732}, with about 1,000 crowd-sourced Python problems for entry-level programmers, covers programming fundamentals and standard library functionality, including task descriptions, code solutions, and three automated test cases for each problem. We also collected the evaluation results for their upgraded versions HumanEval+~\citep{LiuXW023} and MBPP+~\citep{LiuXW023}. Please refer to Appendix~\ref{subapd:main} for details.


\noindent\textbf{Evaluation Metrics}:
%
We use \textbf{EM@$k$} for token-level generation: For this metric, we generate $n \geq k$ samples per instance (with $n=100$ and $k\in\{1, 3, 10\}$ for our experiments). We count the number of correct samples $c \leq n$ judged by exact matching. @$k$ is defined as the average performance over the task, calculated as $\mathbb{E} \left[ 1 - \frac{{\binom{n-c}{k}}} {\binom{n}{k}} \right]$, which is the same with Pass@$k$~\citep{abs-2107-03374}.

%

\subsection{Results and Analysis}
\label{subsect:token_results}

\begin{figure*}[t]
    \includegraphics[width=1.0\linewidth, bb=0 0 960 250]{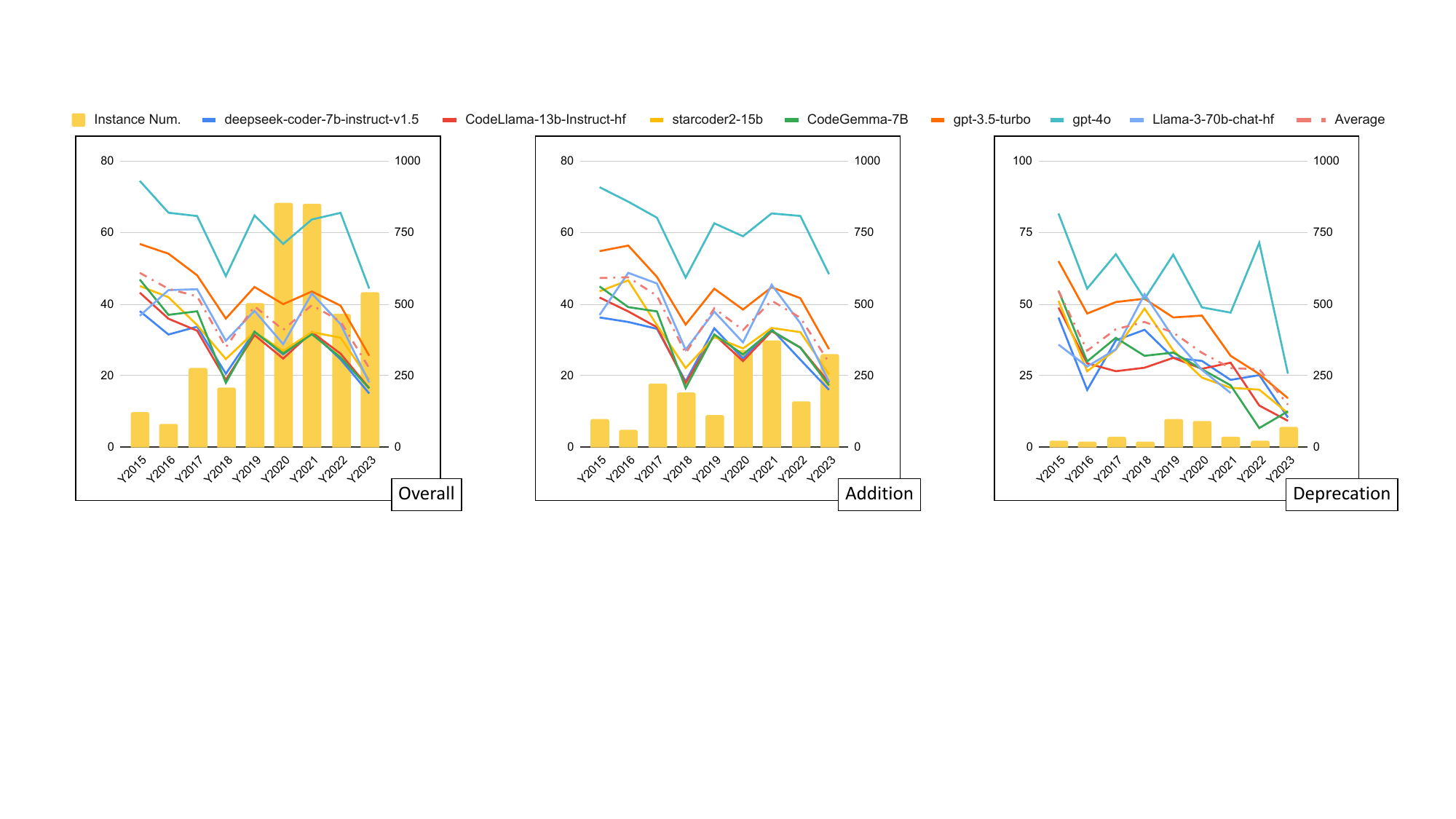}
    \caption{{\small The \emph{EM@1} results for token-level code completion from VersiCode: (a1) Comparison with existing benchmark datasets, (a2) Performance grouped by data sources, and (b) Performance grouped by API lifecycle.\label{fig:main}}}
\end{figure*}

\noindent\textbf{Even token-level code completion is challenging.}
We present the EM@1 results for token-level code completion on VersiCode, sorted by release time (highlighted in green, see Figure~\ref{fig:main}-a1).
Compared to the Pass@1 results on HumanEval (blue) and MBPP (orange), all models perform significantly worse on VersiCode (green). This result indicates the difficulty in disambiguating and recalling version-specific library usage. It is important to note that larger and more recent models, such as GPT-4o (M13) and LLaMA3-70B (M12), demonstrate significantly superior performance compared to other models.
(See Appendix~\ref{apd:negative_example} for the error analysis of GPT-4o.)
However, a substantial performance gap of at least 15 points remains when compared to HumanEval and MBPP (detailed in Appendix~\ref{subapd:main}). This indicates that state-of-the-art LLMs still struggle to deliver satisfactory results, even for the simplest token-level completion tasks.

\noindent\textbf{Differences in LLM performance across different data sources.}
Figure~\ref{fig:main}-a2 presents the EM@1 results for token-level code completion on VersiCode, categorized by data sources.
Among the three data sources, most models perform significantly better on Stack Overflow, especially compared to handling source code from downstream applications. This discrepancy may be attributed to the greater diversity found in downstream applications, which demands a more robust capability to address varied challenges. This may also indicate that Stack Overflow is heavily represented in the pre-training data of LLMs, increasing the likelihood of data leakage. Similar to Figure~\ref{fig:main}-a1, GPT-4o (M13) and LLaMA3-70B (M12) stand out as outliers, excelling in handling downstream applications, which may increase the likelihood of models memorizing specific content. Full numeric results are provided in Appendix~\ref{subapd:main}.

\noindent\textbf{Challenges in casual intermediate library versions.}
We present the token-level EM@1 results for the token-level code completion task, categorized by lifespan features: addition (in blue), deprecation (in orange), and general (referring to intermediate versions; in green),  as shown in Figure~\ref{fig:main}-b. Most models perform well in cases of addition and deprecation, likely because newly added or deprecated APIs are often emphasized in documentation and by the community. 
However, most models struggle with reasoning and adapting to intermediate versions.
As shown in Figure~\ref{fig:main}-a2, models like LLaMA3-70B excel in downstream applications and handle intermediate versions more effectively, likely due to the diversity of use cases they encounter.

\begin{figure}[t]
    \centering
    \includegraphics[width=.98\textwidth,bb=0 0 945 300]{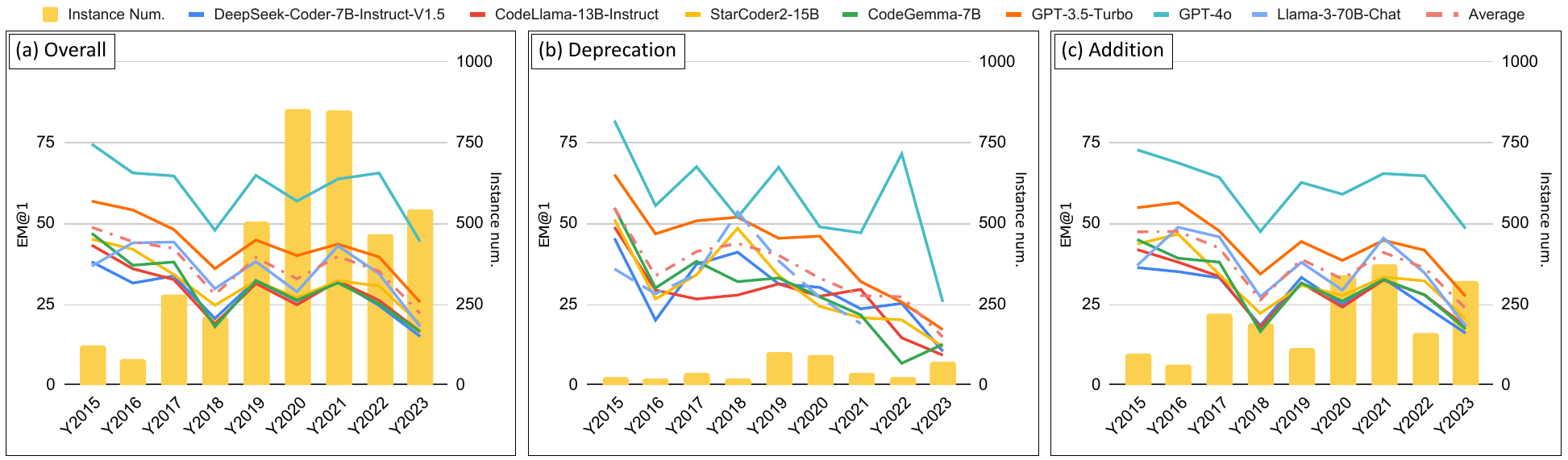}
    \caption{{\small The \emph{EM@1} performance for token-level code completion, grouped by year (2015-2023), with a histogram of data distribution for each year.\label{fig:time_analysis}}}
\end{figure}

\noindent\textbf{The programming knowledge of LLMs, particularly regarding version-specific information, is surprisingly outdated.}
Figure~\ref{fig:time_analysis} presents the EM@1 performance for token-level code completion, grouped by year from 2015 to 2023, along with a histogram showing the data distribution for each year.
To ensure precise timestamps and minimize noise, we only used instances collected from library source code.
As shown in Figure~\ref{fig:time_analysis}-a, there is a clear trend: model performance declines as the release time becomes more recent.
This is counter-intuitive compared to temporal knowledge question answering~\citep{abs-2402-16797}, where performance initially increases before declining. We further filtered for ``deprecation'' (Figure~\ref{fig:time_analysis}-b) and ``addition'' (Figure~\ref{fig:time_analysis}-c) to identify version-sensitive cases. 
Although data sparsity reduces confidence in the results, both cases show a clear downward trend over time
This suggests that LLMs have outdated programming knowledge, highlighting the need for rapid adaptation to newer libraries and APIs.

%% file: sections/3_2_block_level_completion.tex
\section{From Token-level to Line- and Block-level Completion}
\label{sect:line_level_completion}

When utilizing third-party code library APIs, LLMs should handle not only API name generation but also parameter preparation and contextual code integration. In this section, we extend the task to line-level (completing a single line) and block-level (completing multiple lines) code generation. This expanded scope presents new challenges for both the model's capabilities and the evaluation methodologies.
(1) How does increasing complexity in line- or block-level code completion affect the LLMs to handle API usage and parameters?
(2) How does having more context (like import statements and specified library version) improve the accuracy of line- and block-level code generation?
(3) Which evaluation metrics best capture the accuracy of line- and block-level code generation, and which is most reliable?

\subsection{Experiment Setup}

\noindent\textbf{Models}:
We selected GPT-4o, GPT-3.5, and LLaMA3 70B, the three models that perform best on token-level code completion, to conduct experiments on line-level or block-level code completion.

\noindent\textbf{Data Preparation}: 
We sample a subset from VersiCode for dynamic code analysis with executable test cases from library source code, focusing on code snippets with complete context (e.g., import statements). GPT-4 was used to refactor the snippets into task functions, followed by test case generation and validation in a version-specific environment. All of the test cases have been manually verified to ensure their correctness. The code completion tasks are categorized into token, line, and block levels. The test cases include return type, normal input, boundary values, and functionality checks (see Appendix~\ref{subapd:data_prepare} for details).

\begin{figure*}
    \centering
    \includegraphics[width=\linewidth]{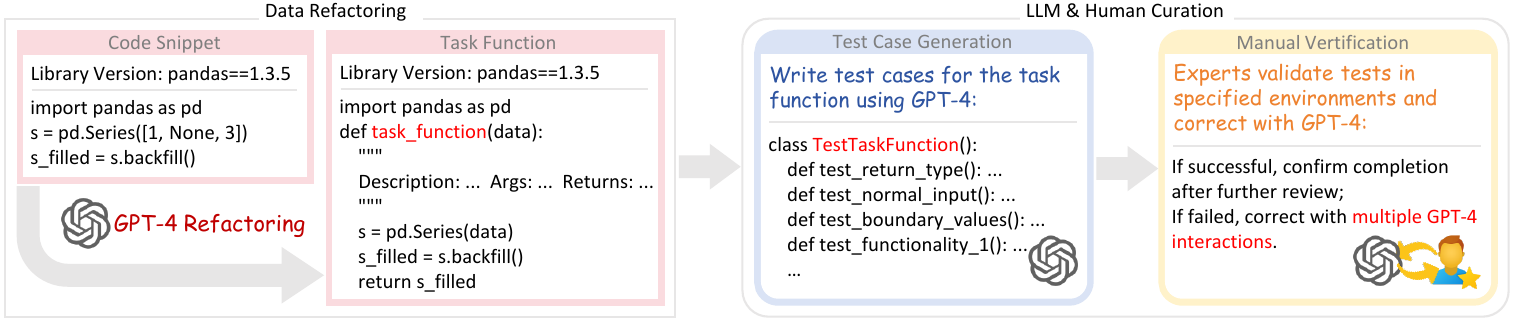}
    \caption{{\small
    The process of executable code assessment, which includes data refactoring, test case generation, and validation. Starting from code snippets collected from real code involving specific API calls for a given library version, GPT-4 is employed to refactor the code into a task function. The large language model is then prompted to generate test cases from various perspectives (See Appendix~\ref{apd:example_execute} for a running example of instances and test cases.) Each generated test case is verified by experts, and the correctness is ensured by running the code in a specified environment. If issues arise, they are corrected through multiple iterations with GPT-4. 
    \label{fig:executable_test_case}
    }}
\end{figure*}

\noindent\textbf{Metrics}:
We use the following evaluation metrics for each task granularity:
(1) \textbf{Pass@$k$} for token-level generation~\citep{abs-2107-03374}:
For this metric, we generate $n \geq k$ samples per instance (with $n=6$ and $k=1$ to compare different metrics). We count the number of correct samples $c \leq n$ judged by executable testing. 
%
(2) \textbf{Identifier Sequence Match (ISM@$k$)} and \textbf{Prefix Match (PM@$k$)} for line-level generation~\citep{abs-2306-10763}: These metrics measure how closely the generated sequences match the ground truth. For block-level generation, we adopt the average performance over lines. Following the setup in Agrawal et al.~\citep{abs-2306-10763}, we generate $n=6$ independent samples per instance. 
(3) \textbf{Exact Match (EM@$k$)}: We use regular expression matching to determine whether the specified API is used in the code generated by the model and the formula for calculating the EM@k score is the same as the formula for calculating the Pass@k score ($n=6$ and $k=1$).
(4) \textbf{Critical Diff Check (CDC@$k$)}: Unlike traditional code similarity calculations, CDC focuses on the differences between the code generated by the model and the reference answer. CDC extends the EM metric by adding four additional rules: checking whether the generated code is syntactically valid; identifying the line in the generated code where the specified API is used and determining if the number of parameters in the function call is the same; if the answer uses a with statement, checking whether the generated code also uses a with statement; and if the answer uses keyword arguments, verifying whether the generated code uses the same keyword arguments.  Please refer to Appendix~\ref{apd:metric_design} for detailed examples, effectiveness analysis, and ablation study, conducted to validate CDC.

\subsection{Results and Analysis}
\label{subsect:block_analysis}

\begin{figure*}
    \centering
    \includegraphics[width=.95\linewidth]{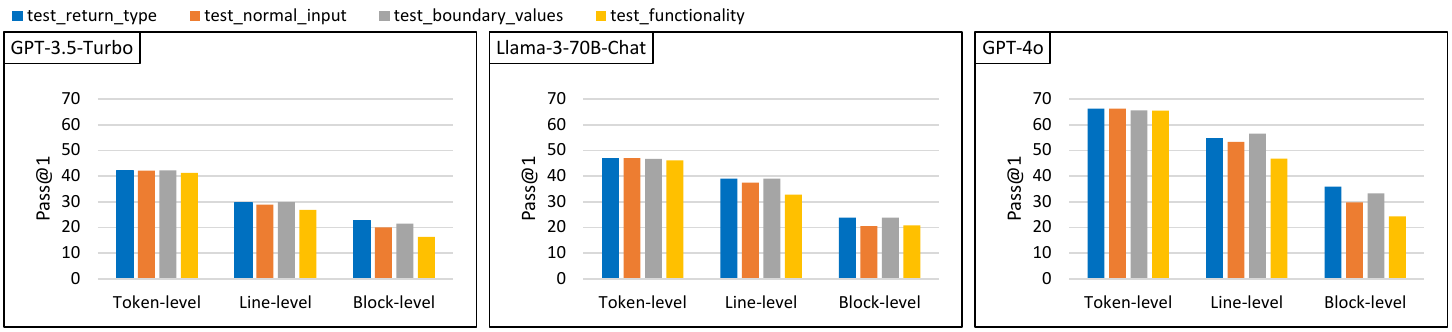}
    \caption{{\small The \emph{Pass@1} performance of different models across various granularities and test case types. }}
    \label{fig:test_cases_performance}
\end{figure*}

\input{table/line_block_completion}


\noindent\textbf{Less context leads to more errors in code generation.}
When models have more context, like import statements, their performance improves significantly. For example, as shown in Table~\ref{tab:metric_gran_completion}, GPT-4o at the token level achieves a Pass@1 score of 65.97 with imports, but this drops to 44.54 without imports. This pattern is consistent across all models and granularity levels (i.e., token, line, and block), as shown in Figure~\ref{fig:test_cases_performance}. When models lack important context, such as external libraries or other dependencies, they struggle to generate accurate code, which leads to more errors. So, giving models more information upfront is crucial for better results.

\noindent\textbf{Models show limited sensitivity to version-specified instructions.} As shown in Table~\ref{tab:metric_gran_completion}, at the token level, models like GPT-4o perform slightly better when provided with version information (52.80 with version v.s. 49.72 without version). However, this advantage diminishes at the line and block levels, where the results become inconsistent. This suggests that while version details can be helpful for short code snippets, they don’t significantly impact the model’s performance for more extended or complex code. This likely indicates that models are not trained to prioritize or heavily rely on version-specific instructions.


\noindent\textbf{The CDC@1 metric closely aligns with Pass@1 scores, making it a strong proxy for dynamic code analysis.} As shown in Table~\ref{tab:metric_gran_completion}, at the block level, the Pearson Correlation Coefficient (PCC) between CDC@1 and Pass@1 is 0.9995, indicating a strong correlation. Even though EM@1 has a high correlation with Pass@1 at the token level (PCC = 0.9995), EM@1 becomes less aligned at the block level (PCC = 0.8974). Additionally, the absolute differences between CDC@1 and Pass@1 values are generally smaller compared to other static metrics like EM@1, making CDC a potentially more reliable alternative for assessing code generation accuracy.

%% file: table/line_block_completion.tex
\begin{table*}[t]
\centering

\includegraphics[width=\linewidth]{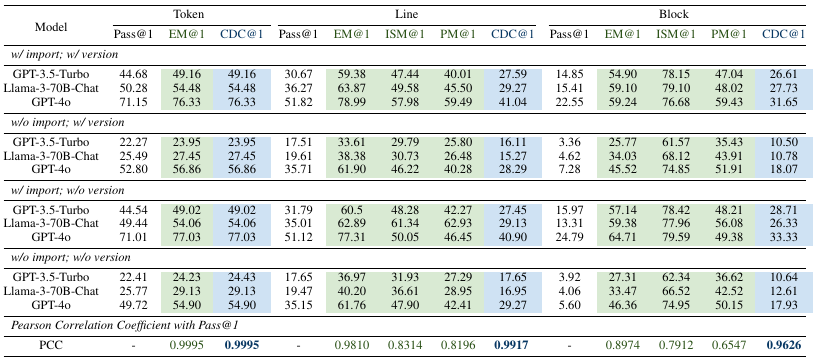}
\caption{\small The performance of different models across various granularities \emph{(Token, Line, Block)}. \emph{Pass@1} refers to dynamic analysis metrics, while green-colored metrics \emph{(EM, ISM, PM)} correspond to static analysis based on string matching. The blue-colored metric \emph{(CDC)} represents a newly proposed metric. The configurations labeled as \emph{``w/o version''} indicate that the prompt does not specify the version of the third-party code libraries, while \emph{``w/o import''} refers to prompts where the provided code context lacks import statements, meaning the model must generate code based entirely on user intent. The Pearson correlation coefficient is computed for each metric's results against Pass@1 within each granularity.}
\label{tab:metric_gran_completion}

\end{table*}

%% file: sections/3_3_migration.tex
\section{From Code Completion to Code Migration}
\label{sect:code_migration}

In addition to generating code for specific third-party library versions, another common challenge is maintaining user projects when these libraries are upgraded or rolled back. 
We address version-aware code migration by exploring three key questions: 
(1) How well can LLMs handle migrating code across different versions, compared to generating code for a specific version? 
(2) What impact do major and minor version changes in third-party libraries have on code migration? 
(3) How do forward migrations (from older to newer versions) compare to reverse migrations (from newer to older versions) in terms of trends and challenges?

\subsection{Experiment Setup}
\noindent\textbf{Models}: Based on the token-level code completion experimental results in Section 3, we selected the most outstanding performers from each model series for the experiments in this section.

\noindent\textbf{Data Preparation}: For code migration, we utilize a subset of VersiCode, in which instances are constructed based on differences between source and target code versions, covering both updates to newer versions and downgrades. Versions were categorized by patterns (e.g., major vs. minor) to capture different migration scenarios. (Detailed in Appendix~\ref{subapd:data_prepare})


\noindent\textbf{Metrics}: Code migration is similar to block-level tasks in code completion. We use the same evaluation metric as for block-level: CDC@$k$ ($n=6$, $k\in\{1, 3\}$).

\subsection{Results and Analysis}
\input{table/merged_migration}

\noindent\textbf{Model performance across version migrations.}
Different models display varying degrees of adaptability when transitioning between major and minor software versions, with some showing exceptional robustness in Table~\ref{tab:migration_results}. The table categorizes version migrations into four types: Major-to-Major, Major-to-Minor, Minor-to-Major, and Minor-to-Minor. Notably, models like GPT-4o excel in major-to-major migrations, suggesting superior handling of significant changes, whereas transitions involving minor versions tend to show moderate performance. This variability underscores the models' design focus, whether on broad adaptability or specialized functionality.

\noindent\textbf{Adaptability in code migration based on release timing.}
Backward and forward compatibility testing reveals a spectrum of model resilience under different temporal migration scenarios. The evaluation is split into two releasing time directions: Old-to-New and New-to-Old, shown in Table~\ref{tab:migration_results}. Generally, models perform better when adapting to newer versions from older ones, with GPT-4o standing out for its high scores in both directions. However, the drop in performance when handling older versions after training on newer releases highlights challenges in maintaining backward compatibility, a critical aspect for long-term usability and integration stability in evolving tech environments.

\noindent\textbf{The context code in another version is still helpful, but its benefits are limited.}
The comparison between block-level code completion and block-level code migration is shown in Table~\ref{tab:migration_results} and Table~\ref{tab:metric_gran_completion}. There is a significant improvement across most models, except for LLaMA3-70B and GPT-4o, detailed in Appendix~\ref{subapd:wo_verify}. When provided with code in another version as context (i.e.\ in the code migration task), these models can generate correct code with a much higher success rate. However, a bottleneck is evident in LLaMA3-70B and GPT-4o, where the code context hinders their performance than code completion.

%% file: table/merged_migration.tex
\begin{table*}[t]
\centering
\includegraphics[width=\linewidth]{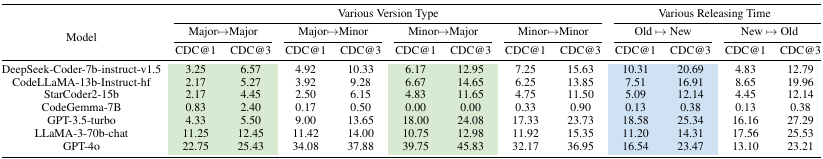}
\caption{{\small
The performance of various models in different code migration scenarios. The arrow "$\mapsto$" indicates the direction of migration, where "Major" corresponds to major version changes (e.g., Torch 2.0.0), and "Minor" corresponds to minor version changes (e.g., Torch 2.1.3). The "Old $\mapsto$ New" scenario simulates upgrading from an old version to a new version, while "New $\mapsto$ Old" represents the maintenance of historical code. The performance of different models in these scenarios is measured using the CDC metrics (CDC@1 and CDC@3), reflecting their adaptability to various code migration tasks. }\label{tab:migration_results}
}
\end{table*}

%% file: sections/4_discussion.tex
\section{Discussion}
\label{sect:discuss}

\noindent\textbf{How can we enhance pre-training for new code-LLMs?}
Figure~\ref{fig:time_analysis} demonstrates a notable decline in the performance of all models over time. This deterioration is likely attributable to two primary factors: (1) the use of outdated pre-training data, which causes older versions of code to predominate the training set, and (2) the backward compatibility of APIs, which results in a higher prevalence of use cases and examples about older versions of these APIs~\citep{LamotheGS22}. To mitigate this issue and improve the models' capabilities with newer libraries, we suggest increasing the representation of new-version codebases within the training data. This adjustment aims to enhance the proficiency in utilizing contemporary libraries effectively~\citep{abs-2402-16797,abs-2402-14526}.
Besides, based  on the results in Section~4.2, current LLMs show limited use of version information in code generation. To address this, we propose enhancing pre-training by incorporating version-tagged code samples and metadata to help models better differentiate between API versions.

\noindent\textbf{How can we address the challenge of evolving libraries in LLMs?}
Generating block-level or repository-level code~\citep{abs-2402-16667} requires LLMs to understand user demands and library dependencies. Addressing this challenge involves continually training the model with new libraries using continual learning techniques~\citep{jiang2024survey}. These techniques enable the model to adapt to changing libraries without forgetting previously learned information. Examples include memory-based methods and various continual learning strategies~\citep{abs-2402-01364, yadav-etal-2023-exploring,WuCLLQH22}. Additionally, developing benchmark datasets that are continuously and automatically curated and maintained is crucial for evaluating the performance of models with new libraries~\citep{JangYLYSHKS22}. Enriching the taxonomy~\citep{JiaoYLQGS23} and maintaining datasets for evolving libraries~\citep{LamotheGS22} is also vital~\citep{JiaoYLQGS23}. Multi-agent systems can be employed for this purpose. 
Aligning development and evaluation efforts will enhance the ability of LLMs in code understanding and generation capabilities, to remain effective as libraries evolve.

\noindent\textbf{Can we address version-controllable code generation with retrieval-augmented generation?}
Retrieval-augmented generation (RAG) approaches typically involve two crucial components: retrieval and in-context generation~\citep{abs-2312-10997}. The following challenges need to be addressed in order for RAG to be effectively applied to this problem. 
From the \textbf{retrieval} perspective:
(1) It may be difficult to disambiguate version-related queries, as embeddings for version strings like ``torch 2.1.3'' and ``torch 1.3.2'' can be very similar~\citep{abs-2402-14903}. This similarity makes it hard for retrievers to differentiate between specific features and capabilities associated with each version. 
( 2) Version information of code snippets is rarely explicitly mentioned within the code itself and may instead appear in separate configuration files like ``requirements.txt''. This separation necessitates a more sophisticated retrieval approach, where the model must integrate information from multiple sources to accurately understand version dependencies.
From the perspective of \textbf{in-context generation}:
Table~\ref{tab:wogrammar} shows that even non-matching version contexts (i.e., code migration) can help smaller models generate grammatically correct code. 
%
This observation suggests potential for dedicated RAG approaches~\citep{jiang2024survey}, though the benefits are limited and retrieval noise may reduce effectiveness.

\noindent\textbf{What are the effective methods for evaluating the capabilities of LLMs in generating version-controllable code? }
Both static analysis~\citep{abs-2306-10763}, which reviews code without executing it, and dynamic analysis~\citep{abs-2406-15877}, which tests the code by running it, are vital for software development. 
However, evaluating LLMs for version-controllable code generation presents unique challenges. 
(1) Dynamic analysis is complicated by API calls that rely on specific code contexts, making it difficult and costly to create standalone tests~\citep{abs-2406-15877}. 
Additionally, using LLM-generated code as test cases introduces further complexity in managing test quality. 
Especially, VersiCode, which includes 300 packages and over 2,000 versions in the raw dataset, requires detailed setups for each testing environment and managing various dependencies, complicating the practical deployment of solutions.  
(2) Meanwhile, static analysis uses metrics like ISM~\citep{abs-2306-10763} and PM~\citep{abs-2306-10763} for broad coverage but may miss critical details such as indentation and parameter positioning in API-related code, refer to Table~\ref{tab:metric_gran_completion} and Appendix~\ref{apd:metric_design}. 
These omissions suggest that traditional static metrics are not entirely suitable for assessing version-controllable code generation.
Evaluating the effectiveness of these metrics is crucial. 
Our study initiates the exploration of more reliable methods; however, extensive research, including approaches like code slicing~\citep{abs-2405-02355}, is essential to advance our evaluation techniques.

%% file: sections/5_related_work.tex
\section{Related Work}
\label{sect:related_work}

\noindent\textbf{Code Generation Models}: 
Recent advancements in code language models~\citep{abs-2401-14196,codegemma_2024,qwen,abs-2308-12950,abs-2403-14734}, driven by sophisticated NLP techniques~\citep{jiang2024survey} and extensive code repositories~\citep{abs-2310-20329}, have resulted in substantial breakthroughs. Transformer-based large language models~\citep{abs-2306-08568,abs-2308-12950,abs-2401-14196,abs-2402-19173,qwen,abs-2306-11644,abs-2309-05463} have demonstrated exceptional capabilities in generating syntactically correct and semantically meaningful code from natural language descriptions. Additionally, research efforts that integrate multi-modal data~\citep{gpt35,gpt4o,llama3modelcard}, including both code and accompanying documentation~\citep{abs-2310-20329}, have significantly improved model accuracy. While in real-world software engineering, 

\noindent\textbf{Code Generation Datasets}: 
The code generation~\citep{jiang2024survey,abs-2403-14734,abs-2403-00338} includes tasks for both code completion and code editing, ensuring comprehensive coverage of programming scenarios.
Code completion~\citep{YaoWCS18,YinDCVN08,FengGTDFGS0LJZ20,abs-2107-03374,abs-2108-07732,HendrycksBKMAGB21,LuGRHSBCDJTLZSZ21,abs-2203-07814,FriedAL0WSZYZL23,LiuXW023,Lai0WZZZYFWY23,YuSRZZMLLWX24,abs-2309-01940,abs-2303-17568} is the task of predicting subsequent code tokens based on the given context, benefits from datasets, which provide extensive code repositories from various programming languages. These datasets enable models to learn syntactic and semantic patterns~\citep{JiaoYLQGS23}. 
%
Code editing~\citep{JustJE14,LinKCS17,Zhu0R22,abs-2206-08474,abs-2310-20329,YanTLCW23,AhmadTCC23,JiaoYLQGS23,abs-2310-08879,abs-2401-04621} involves automatically generating changes to existing code, such as bug fixes or refactoring. Datasets like EvalGPTFix~\citep{abs-2310-08879} and DebugBench~\citep{abs-2401-04621}, which focus on bug fixing and code refinement tasks, are instrumental in this area. 
%
%
To our knowledge, given the necessity and challenges in library evolution~\citep{jiang2024survey}, refer to the detailed comparison in Table~\ref{tab:com_edit} and Appendix~\ref{subapd:completion}, the proposed dataset VersiCode is the first large-scale code generation dataset, covering both code completion and code editing. Refer to Appendix~\ref{apd:related_data} for a comprehensive comparison among datasets.

\noindent\textbf{Third-party Library Evolution: } Third-party library code is continually updated due to bug fixes, code refactoring, and the addition of new features, making it a significant research topic in software engineering~\citep{ZhangZWTLX20,ZhangY00RG21,DilharaKD21,Liu0YFG21,Wang0LC20,VadlamaniKC21,HaryonoT0LJ21}. 
Studies by \cite{ZhangZWTLX20} show that Python APIs often evolve by adding, deleting, or modifying parameters. Further research by \cite{ZhangY00RG21} notes frequent API changes, including parameter updates. 
\cite{DilharaKD21} reveal that developers adjust their use of machine learning libraries in response to updates, while \cite{Liu0YFG21} and \cite{DigJ06} find that undocumented changes in Android and Java can cause errors.
Research on API deprecation highlights issues with documentation and the quality of suggested alternatives \citep{Wang0LC20,VadlamaniKC21,HaryonoT0LJ21,BritoHVR18}, showing that improvement in library evolution does not necessarily translate to better suggestions for deprecated APIs.
VersiCode, unlike traditional software engineering research, studies API version evolution from an LLM perspective, exploring its impact on model training, code generation, and evaluation.

%% file: sections/6_conclusion.tex
\section{Conclusion}

In conclusion, our research underscores the need for updated benchmarks that capture the dynamic nature of software development, better assessing the capabilities of LLMs in code generation. By introducing the VersiCode dataset, we provide a realistic testing ground that reveals significant limitations in current models, like GPT-4o and LLaMA3, when handling version-specific code. Our findings advocate for continuous model improvements and the adoption of our new metric, i.e., critical diff check, which more accurately evaluates model performance against real-world challenges. This work not only introduces valuable tools but also sets a direction for future enhancements in AI-driven code generation, ensuring LLMs remain effective and relevant in professional settings.
For future research, we will investigate a solution for version-controllable code generation based on the insights from this paper, including approaches like continual learning, memory-enhanced methods, or retrieval-based methods. Additionally, we plan to develop a live version of VersiCode, which will continuously incorporate new libraries and downstream use cases.

%% file: sections/A_dataset_construction.tex
\section{Dataset Construction}
\label{apd:dataset_construct}

VersiCode is a large-scale code generation benchmark dataset focusing on evolving library dependencies. We propose two tasks to simulate real-world applications: version-specific code completion and version-aware code migration, incorporating version information into code generation constraints.
First, we discuss data curation, and preprocessing of noisy code snippets and FAQs into organized metadata. Based on the metadata, we describe the task design and quality control process. We then address tagging API lifespan features per library version. Finally, we provide data statistics for VersiCode and discuss future dataset extensions.

\subsection{Dataset Curation and Collection}
\label{subapd:data_curation}
\begin{figure*}[t]
    \centering
    \includegraphics[width=1.0\linewidth, bb=0 0 645 210]{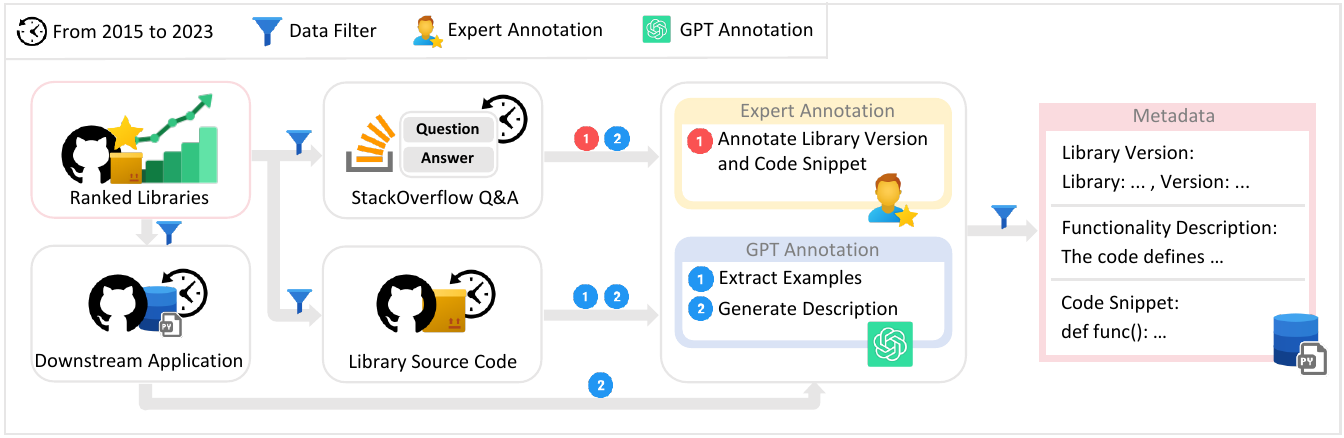}
    \caption{{\small The preprocessing pipeline to obtain metadata, structured as n-gram tuple of $\langle$library name, version, functionality description, code snippet$\rangle$.\label{fig:preprocess}}}
\end{figure*}

As shown in Figure~\ref{fig:preprocess}, we first collected permissively licensed Python repositories from GitHub that serve as the source code for Python libraries. These repositories are ranked by their popularity (as indicated by their collected stars). Using the list of popular libraries, we gathered data from three sources for each library:
(1) Library Source Code: We collected all available versions of the library source code from GitHub, verifying with PyPI to ensure that the collected versions are formally released and can be installed via pip. 
From the library source code, we extracted official usage examples for each API from the docstrings.
(2) Downstream Application Code: Given Python's popularity in scientific programming, we collected the source code from top-tier research papers over 10 years as downstream applications. These applications are valuable due to being lightweight yet self-consistent, diverse in their topics, and tagged release timelines associated with publishing venues. Given the time span, this data source implicitly includes evolving libraries.
(3) Stack Overflow: Using the library names as queries, we collected FAQ data from Stack Overflow, which provides real user queries and diverse user answers. We filtered the data to include only those queries that explicitly mention the versions of the libraries used, using heuristic rules, as shown in Table~\ref{tab:rule}. Additionally, 
we have made our best efforts to filter all of the source code based on the open-source licenses of the repositories to ensure there is no infringement.

\input{table/annotation_rule}

Given the high diversity and varied quality of the collected raw data, we adopted a hybrid annotation approach involving both human experts and LLMs, such as ChatGPT. (1) Library Source Code: The library version is concrete and explicitly available, but example usage varies across libraries and versions. We used an LLM with in-context learning to help extract example code from docstrings, preparing the library version and code snippets. (2) Downstream Applications: The version can easily be extracted from configuration files, typically named ``requirements.txt''. We carefully filtered out Python files that are too long, do not mention the library version, or fail to compile. (3) Stack Overflow: Given the diversity of the questions, we designed strict heuristic rules to preliminarily annotate the library name, version, and corresponding Python code snippets mentioned in answers. We then distributed the pre-annotated data to six qualified human experts for verification and correction, ensuring the library version and code snippets are ready as well.
With all pairs of library versions and code snippets, we employed ChatGPT with in-context learning to generate descriptions of the functionality for each code snippet. Each pair is wrapped in well-organized metadata.

\subsection{Lifecycle Tagging of APIs}
\label{subapd:back_compatib}
Consider an API $a$ added to the library $L$ in version $V_s$ and deprecated in version $V_e$, and is active in the intermediate version $V_m$ where $s \leq m \leq e$. We refer to the interval $[s,e)$ as the \emph{lifespan} of $a$. To analyze model performance in detail, we assessed how up-to-date each LLM was concerning newly added or deprecated APIs per version. We compared the source code between any two consecutive versions of each library to detect changes in API or method names. Based on the detection results, we labeled the datasets obtained from the library source code as follows: ``addition'' indicates an API newly added in the current version and still applicable in subsequent versions; ``deprecation'' indicates the current version is the last usable version for the API; and ``general'' indicates the API usage method is inherited from the previous version.

\subsection{Data Preparation for Evaluation}
\label{subapd:data_prepare}

\noindent\textbf{Data Preparation for Token-level Code Completion.} As introduced in Section~\ref{sect:task_definition}, we designed two types of version-controllable code generation tasks: version-specific code completion and version-aware code migration. The task granularities are categorized into token-level, line-level, and block-level to control difficulty and simulate different application scenarios. To better understand model performance, each instance in VersiCode is also tagged with the following: 
(1) Data source, which includes library source code, downstream applications, and Stack Overflow; 
(2) Feature type, including addition, deprecation, and general; 
(3) Release time, i.e.\ the timestamp from GitHub and Stack Overflow); 
%
%
These tags allow us to filter the evaluation dataset and gain sharper insights into model performance.

\noindent\textbf{Data Preparation for Execution-based Multi-granularity Code Completion.}
As shown in Figure~\ref{fig:executable_test_case}, we have constructed a subset for dynamic code analysis that includes executable test cases. From the data originating from library source code in VersiCode, we filter for data that includes complete context (e.g., import statements) code snippets. Experts interact with the web version of GPT-4 to refactor the code snippets into task functions. After a manual check of the task functions, experts interact with GPT-4 to write test cases for them. During the interaction, experts provide appropriate feedback to GPT-4. The test cases are run in a testing environment containing specific library versions (e.g., pandas==1.3.5); if successful, the annotation is completed after further manual verification, and if failed, more detailed feedback is provided to GPT-4 to assist with corrections. The annotated task function is processed into code completion forms with three levels of $\langle$mask$\rangle$ granularity: token, line, and block.
The executable test cases include four types: 
(1) \textbf{Test return type}: tests whether the return type is correct. 
(2) \textbf{Test normal input}: tests whether the expected output is produced with normal inputs. 
(3) \textbf{Test boundary values}: tests whether special values (such as null values, incorrect types, etc.) are handled properly.
(4) \textbf{Test functionality}: tests whether the function fulfills its primary functionality. The first three types of test cases have one instance per task function, while the fourth type has 1-3 instances.

\noindent\textbf{Data Preparation for Code Migration.} As shown in Figure~\ref{fig:postprocess}, considering code migration instances constructed from pairs of metadata, the differences between source and target code versions result in various situations, such as updates from an older version to a newer version or vice versa. Additionally, we categorized versions according to version patterns, for example, treating torch v1.0.0 as a major version and torch v1.3.1 as a minor version, to identify combinations of major and minor version migration cases.

%% file: table/annotation_rule.tex
\begin{table}[t]
\centering
\resizebox{1.0\textwidth}{!}{
\begin{tabular}{@{}cl@{}}
\toprule
\textbf{Procedure}                                             & \textbf{Rules}                                                                                                                                                                                                                                                                                                                                                                                                           \\ \midrule
Ranked Libraries $\mapsto$ StackOverflow Q\&A     & \begin{tabular}[c]{@{}l@{}}Filter out answers that involve the use of libraries from the ranked libraries, \\ and ensure these answers include content in the library version format (e.g., pandas==1.3.5) as well as code snippets.\end{tabular}                                                                                                                                                               \\ \midrule
Ranked Libraries $\mapsto$ Library Source Code    & Based on the ranked libraries, parse the source code of these libraries to find functions related to version changes.                                                                                                                                                                                                                                                                                           \\ \midrule
Ranked Libraries $\mapsto$ Downstream Application & \begin{tabular}[c]{@{}l@{}}(1)Exclude files that do not utilize libraries and version information explicitly listed in requirements.txt.\\ (2)Exclude files with an average line length exceeding 100 characters.\\ (3)Exclude files with a maximum line length exceeding 1000 characters.\\ (4)Exclude files with less than 25\% of alphabetic characters.\\ (5)Exclude files with syntax errors.\end{tabular} \\ \midrule
\multirow{3}{*}{Annotation $\mapsto$ Metadata}    & \begin{tabular}[c]{@{}l@{}}StackOverflow: Filter out data that has been annotated by experts with correct library version and code snippet, \\ and utilize GPT to generate functionality descriptions for the code snippets.\end{tabular}                                                                                                                                                                       \\ \cmidrule(l){2-2} 
                                                      & \begin{tabular}[c]{@{}l@{}}Library Source Code: Utilize GPT to extract examples from version change function docstrings, \\ filter out successfully extracted data, and employ GPT to generate functionality descriptions for the examples.\end{tabular}                                                                                                                                                        \\ \cmidrule(l){2-2} 
                                                      & Downstream Application: Utilize GPT to generate functionality descriptions for code snippets.                                                                                                                                                                                                                                                                                                                   \\ \bottomrule
\end{tabular}}
\caption{{\small Detailed explanation of annotation stages and the corresponding filtering rules.\label{tab:rule}}}
\end{table}

%% file: sections/B_1_dataset_statistics.tex
\section{Data Statistics and Scope}
\label{apd:data_statistics}

\noindent\textbf{Dataset Statistics}: We present the statistics of VersiCode in Table~\ref{tab:data_statist}, using the StarCoder2's~\citep{abs-2402-19173} tokenizer to compute the number of tokens. We also outline the complete version of VersiCode in the table, which furnishes human-labeled data for three additional languages: C\#, Java, and JavaScript. Our executable data, applied in Section~\ref{sect:line_level_completion}, is a high-quality human-annotated subset from VersiCode, covering 12 libraries, 40 versions, and 119 functionality descriptions. For each functionality description, we matched 4 to 5 test cases.

\input{table/data_statis}

\begin{figure*}[t]
    \centering
    \includegraphics[width=\linewidth]{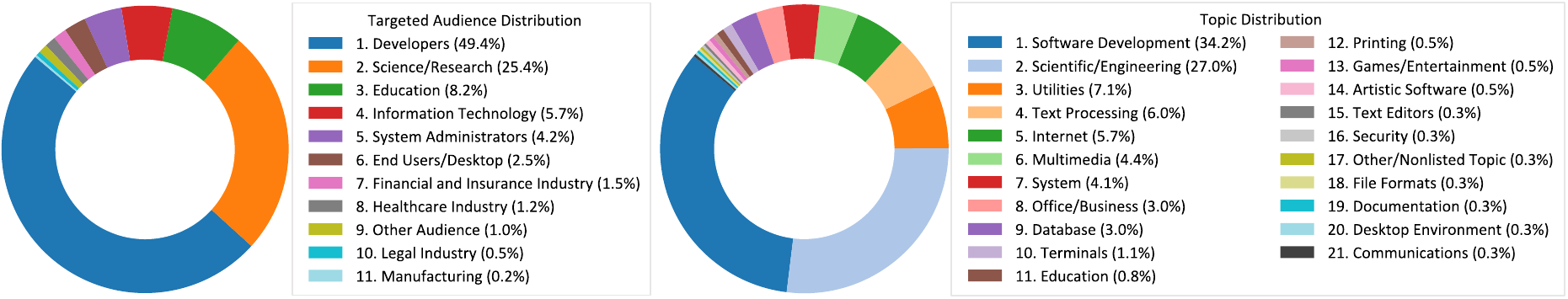}
    \caption{{\small A proportional chart based on the classification system of targeted audience and topics in third-party Python libraries on PyPI.}}
    \label{fig:topic_audience_all}
\end{figure*}

\noindent\textbf{Scope}: VersiCode supports version-specific code completion at the token, line, and block levels, enabling developers to navigate through version variations effortlessly. It also facilitates block-level version-aware code editing, empowering users to make precise modifications tailored to requirements of each version. 
The collected metadata also serves as a valuable resource for potential customized task modifications, supported domains are illustrated in Figure~\ref{fig:topic_audience_all}, aiding in fine-tuning workflows and enhancing model training for optimal performance.

%% file: table/data_statis.tex
\begin{table}[tbh]
\centering
\resizebox{\textwidth}{!}{
\begin{tabular}{@{}ccccccccc@{}}
\toprule
        \textbf{\# Language}            & \multicolumn{5}{c}{\textbf{Python} }                                                     & \textbf{Java}          & \textbf{C\#}           & \textbf{JavaScript}    \\ 
                    \cmidrule(r){1-1}\cmidrule(lr){2-6}\cmidrule(lr){7-7}\cmidrule(lr){8-8}\cmidrule(l){9-9}
\# Data Source       & \multicolumn{5}{c}{StackOverflow; Library Source Code; Downstream Application}  & StackOverflow & StackOverflow & StackOverflow \\ 
\midrule
\# Num. of Libraries & \multicolumn{5}{c}{300}                                                         & 19            & 16            & 33            \\ \midrule
\# Num. of Versions  & \multicolumn{5}{c}{2,207}                                                       & 25            & 16            & 60            \\ \midrule
\# Size of Meta Data & \multicolumn{5}{c}{11,268}                                                      & 29            & 16            & 62            \\ \midrule
\# Task Type         & \multicolumn{3}{c}{Completion} & Editing (old to new)   & Editing (new to old)  & Completion    & Completion    & Completion    \\ 
\cmidrule(r){1-1}\cmidrule(lr){2-4}\cmidrule(lr){5-5}\cmidrule(lr){6-6}\cmidrule(lr){7-7}\cmidrule(lr){8-8}\cmidrule(l){9-9}
\# Granularity       & Token     & Line     & Block   & Block                  & Block                 & Block         & Block         & Block         \\ \midrule
\# Avg. Input Token  & 2,087     & 2,075    & 55      & 191                    & 195                   & 57            & 63            & 67            \\ \midrule
\# Avg. Output Token & 2         & 16        & 128      & 131                     & 128                    & 349           & 255           & 167           \\ \midrule
\# Num. of Instances & 13,488    & 13,490   & 1,617   & 38,037                 & 38,037                & 32            & 21            & 82            \\ \bottomrule
\end{tabular}}
\caption{{\small Data statistics of VersiCode, including multiple languages. \label{tab:data_statist}}}
\end{table}

%% file: sections/B_2_related_dataset.tex
\section{Related Dataset}
\label{apd:related_data}

\noindent\textbf{Code Completion Datasets.}
\label{subapd:completion}
As shown in Table~\ref{tab:complet}, we compare the VersiCode-completion dataset with existing benchmarks. VersiCode stands out in annotated data size, marking it as the inaugural dataset tailored for version-specific generation.
\input{table/code_completion_dataset}

\noindent\textbf{Code Migration Datasets.}
\label{subapd:migration}
As shown in Table~\ref{tab:com_edit}, we compare the VersiCode-migration dataset with existing benchmarks. VersiCode stands out in annotated data size, marking it the inaugural dataset tailored for version-specific migration.
\input{table/code_editing_dataset}

%% file: table/code_completion_dataset.tex
\begin{table}[htb]
\centering

\resizebox{1.0\textwidth}{!}{
\begin{tabular}{@{}cccccccc@{}}
\toprule
        \textbf{Benchmark}                                                                              & \textbf{Source}                         & \textbf{Language }                    & \textbf{Samples} & \textbf{Completion Task}                  & \textbf{Granularity }                         & \textbf{Collection Time} & \textbf{Annotation} \\ \midrule
StaQC~\citep{YaoWCS18}                                                                  & StackOverflow                  & Python, SQL                  & 267,056 & Function Programming             & Line-Level, Block-Level              & 2018            & None       \\ \midrule
CoNaLa~\citep{YinDCVN08}                                                                & StackOverflow                  & Python, Java                 & 2,879   & Function Programming             & Line-Level, Block-Level              & 2018            & Human      \\ 
\midrule
CT-maxmin~\citep{FengGTDFGS0LJZ20}                                                      & Existing Benchmark             & Multi(=6)                    & 2,615   & Cloze Test                       & Token-Level                          & 2020            & None       \\ \midrule
HumanEval~\citep{abs-2107-03374}                                                        & Hand-Written                   & Python                       & 164     & Function Programming             & Line-Level, Block-Level              & 2021            & Human      \\ \midrule
MBPP~\citep{abs-2108-07732}                                                             & Hand-Written                   & Python                       & 974     & Function Programming             & Block-Level                          & 2021            & Human      \\ \midrule
APPS~\citep{HendrycksBKMAGB21}                                                          & Programming Sites              & Python                       & 10,000  & Function Programming             & Line-Level, Block-Level              & 2021            & None       \\ \midrule
CT-all~\citep{LuGRHSBCDJTLZSZ21}                                                        & Existing Benchmark             & Multi(=6)                    & 176,115 & Cloze Test                       & Token-Level                          & 2021            & None       \\ \midrule
CodeContests~\citep{abs-2203-07814}                                                     & Existing Benchmark, Codeforces & Multi(=3)                    & 13,610  & Function Programming             & Block-Level                          & 2022            & None       \\ \midrule
HumanEval-FIM~\citep{FriedAL0WSZYZL23}                                                  & Existing Benchmark             & Python                       & 164     & Function Programming             & Line-Level, Block-Level              & 2022            & None       \\ \midrule
HumanEval+~\citep{LiuXW023}                                                             & Existing Benchmark             & Python                       & 164     & Function Programming             & Line-Level, Block-Level              & 2023            & LLM      \\ \midrule
MBPP+~\citep{LiuXW023}                                                                  & Existing Benchmark             & Python                       & 378     & Function Programming             & Block-Level                          & 2023            & LLM      \\ \midrule
DS-1000~\citep{Lai0WZZZYFWY23}                                                          & StackOverflow                  & Python                       & 1,000   & Function Programming             & Line-Level, Block-Level              & 2023            & Human      \\ \midrule
CoderEval~\citep{YuSRZZMLLWX24}                                                         & Github                         & Python, Java                 & 460     & Function Programming             & Block-Level                          & 2023            & Human      \\ \midrule
CodeApex~\citep{abs-2309-01940} & Programming Sites              & C++                          & 476     & Function Programming             & Block-Level                          & 2023            & None       \\ \midrule
HumanEval-X~\citep{abs-2303-17568}                                                      & Existing Benchmark             & Multi(=5)                    & 820     & Function Programming             & Line-Level, Block-Level              & 2023            & Human      \\ \midrule
BigCodeBench~\citep{abs-2406-15877}                                                      & Existing Benchmark             & Python                    & 1,140     & Function Programming             & Block-Level              & 2024            & Human, LLM      \\ \midrule
\textbf{VersiCode}                                                                              & StackOverflow, Github          & Python, Java, C\#, JavaScript & 28,595  & Cloze Test, Function Programming & Token-Level, Line-Level, Block-Level & 2024            & Human, LLM \\ \bottomrule
\end{tabular}
}
\caption{{\small Comparison of VersiCode and other code completion datasets. VersiCode is the largest annotated dataset, covering multiple languages and granularities, and involving both human and LLM joint annotations.\label{tab:complet}}}
\end{table}

%% file: table/code_editing_dataset.tex
\begin{table}[htb]
\centering

\resizebox{1.0\linewidth}{!}{
\begin{tabular}{@{}cccccccc@{}}
\toprule
\textbf{Benchmark} & \textbf{Source} & \textbf{Language} & \textbf{Samples} & \textbf{Editing Task} & \textbf{Granularity} & \textbf{Collection Time} & \textbf{Annotation}                 \\ \midrule
\multicolumn{1}{c}{Defects4J~\citep{JustJE14}}                                                             & \multicolumn{1}{c}{Open Source Programs}              & \multicolumn{1}{c}{Java}         & \multicolumn{1}{c}{357}       & \multicolumn{1}{c}{Debug}            & \multicolumn{1}{c}{Block-Level}              & \multicolumn{1}{c}{2014} & \multicolumn{1}{c}{None}  \\ \midrule
\multicolumn{1}{c}{QuixBugs~\citep{LinKCS17}}                                                              & \multicolumn{1}{c}{Quixey Challenges}                 & \multicolumn{1}{c}{Python, Java} & \multicolumn{1}{c}{40}        & \multicolumn{1}{c}{Debug}            & \multicolumn{1}{c}{Line-Level}               & \multicolumn{1}{c}{2017} & \multicolumn{1}{c}{Human} \\ \midrule
\multicolumn{1}{c}{CoST~\citep{Zhu0R22}}                                                                   & \multicolumn{1}{c}{GeeksForGeeks}                     & \multicolumn{1}{c}{Multi(=7)}    & \multicolumn{1}{c}{132,046}   & \multicolumn{1}{c}{Code Translation} & \multicolumn{1}{c}{Line-Level, Block-Level}  & \multicolumn{1}{c}{2022} & \multicolumn{1}{c}{None}  \\ \midrule
\multicolumn{1}{c}{XLCoST~\citep{abs-2206-08474}} & \multicolumn{1}{c}{GeeksForGeeks}                     & \multicolumn{1}{c}{Multi(=8)}    & \multicolumn{1}{c}{1,083,000} & \multicolumn{1}{c}{Code Translation} & \multicolumn{1}{c}{Line-Level, Block-Level}  & \multicolumn{1}{c}{2022} & \multicolumn{1}{c}{None}  \\ \midrule
\multicolumn{1}{c}{InstructCoder~\citep{abs-2310-20329}}                                                   & \multicolumn{1}{c}{Github}                            & \multicolumn{1}{c}{Python}       & \multicolumn{1}{c}{114,000}   & \multicolumn{1}{c}{Code Refinement}  & \multicolumn{1}{c}{Block-Level}              & \multicolumn{1}{c}{2023} & \multicolumn{1}{c}{LLM}   \\ \midrule
\multicolumn{1}{c}{MultilingualTrans~\citep{YanTLCW23}}                                                    & \multicolumn{1}{c}{Programming Sites}                 & \multicolumn{1}{c}{Multi(=8)}    & \multicolumn{1}{c}{30,419}    & \multicolumn{1}{c}{Code Translation} & \multicolumn{1}{c}{Block-Level}              & \multicolumn{1}{c}{2023} & \multicolumn{1}{c}{None}  \\ \midrule
\multicolumn{1}{c}{NicheTrans~\citep{YanTLCW23}}                                                           & \multicolumn{1}{c}{Programming Sites}                 & \multicolumn{1}{c}{Multi(>8)}    & \multicolumn{1}{c}{236,468}   & \multicolumn{1}{c}{Code Translation} & \multicolumn{1}{c}{Block-Level}              & \multicolumn{1}{c}{2023} & \multicolumn{1}{c}{None}  \\ \midrule
\multicolumn{1}{c}{LLMTrans~\citep{YanTLCW23}}                                                             & \multicolumn{1}{c}{Hand-Written}                      & \multicolumn{1}{c}{Multi(=8)}    & \multicolumn{1}{c}{350}       & \multicolumn{1}{c}{Code Translation} & \multicolumn{1}{c}{Block-Level}              & \multicolumn{1}{c}{2023} & \multicolumn{1}{c}{Human} \\ \midrule
\multicolumn{1}{c}{Avatar~\citep{AhmadTCC23}}                                                              & \multicolumn{1}{c}{Programming Sites}                 & \multicolumn{1}{c}{Python, Java} & \multicolumn{1}{c}{62,520}    & \multicolumn{1}{c}{Code Translation} & \multicolumn{1}{c}{Block-Level}              & \multicolumn{1}{c}{2023} & \multicolumn{1}{c}{None}  \\ \midrule
\multicolumn{1}{c}{G-TransEval~\citep{JiaoYLQGS23}}                                                        & \multicolumn{1}{c}{Existing benchmark, GeeksForGeeks} & \multicolumn{1}{c}{Multi(=5)}    & \multicolumn{1}{c}{400}       & \multicolumn{1}{c}{Code Translation} & \multicolumn{1}{c}{Token-Level, Block-Level} & \multicolumn{1}{c}{2023} & \multicolumn{1}{c}{Human} \\ \midrule
\multicolumn{1}{c}{EvalGPTFix~\citep{abs-2310-08879}}                                                      & \multicolumn{1}{c}{AtCoder}                           & \multicolumn{1}{c}{Java}         & \multicolumn{1}{c}{151}       & \multicolumn{1}{c}{Debug}            & \multicolumn{1}{c}{Block-Level}              & \multicolumn{1}{c}{2023} & \multicolumn{1}{c}{Human} \\ \midrule
\multicolumn{1}{c}{DebugBench~\citep{abs-2401-04621}}                                                      & \multicolumn{1}{c}{LeetCode}                          & \multicolumn{1}{c}{Multi(=3)}    & \multicolumn{1}{c}{4,253}     & \multicolumn{1}{c}{Debug}            & \multicolumn{1}{c}{Block-Level}              & \multicolumn{1}{c}{2024} & \multicolumn{1}{c}{LLM}   \\ \midrule
\textbf{VersiCode}                                                                                                   & Github                                                 & Python                            & 76,074                         & Version Adaptation                    & Block-Level                                   & 2024                      & LLM                        \\ \bottomrule
\end{tabular}
}
\caption{{\small Comparison between VersiCode and other code editing datasets, with VersiCode standing out as the largest annotated dataset specifically tailored for version adaptation.\label{tab:com_edit}}}
\end{table}

%% file: sections/C_additional_experiments.tex
\section{Additional Experiments and Details}
\label{apd:add_exp}

\subsection{Extensive Comparative Study on Large Language Models}
\label{subapd:main}

In addition to the model depicted in Figure~\ref{fig:main}, comprehensive and detailed evaluation results are presented in Table~\ref{tab:main}, encompassing 23 models and sorted by the release time of each model.

\input{table/main}

\subsection{Multi-language Analysis}
\label{subapd:mla}
\input{table/multi_language}

As depicted in Table~\ref{tab:language}, we perform the primary multi-language experiments. Counter-intuitively, the performance of LLMs in Java, JavaScript, and C\# surpasses that in Python. This anomaly might be attributed to potential data leakage from the Stack Overflow dataset.

\subsection{Block-level Generation without Grammar Verification}
\label{subapd:wo_verify}

We use Python's built-in function \emph{``compile()''} to compile the generated code snippets in order to check whether they are syntactically correct. Upon comparing ``w/o grammar verification'' and ``w\ grammar verification'' in Table~\ref{tab:wogrammar}, it becomes evident that the model tasked with editing, alongside reference code snippets from other versions, finds it easier to produce grammar-verified code.

\input{table/wo_grammar_verfication}

%% file: table/main.tex
\begin{table}[t]
\centering
\resizebox{\textwidth
}{!}{
\begin{tabular}{@{}cccccccccc@{}}
\toprule
\multirow{2}{*}{\textbf{Release Time}} & \multirow{2}{*}{\textbf{Model}}                                                     & \textbf{HumanEval} & \textbf{HumanEval+} & \textbf{MBPP}   & \textbf{MBPP+}  & \multicolumn{4}{c}{\textbf{VersiCode}}                                                                                                      \\ \cmidrule(lr){3-4}\cmidrule(lr){5-6} \cmidrule(l){7-10}
                              &                                                                            & EM@1    & EM@1     & EM@1 & EM@1 & \multicolumn{1}{c}{Library Source Code} & \multicolumn{1}{c}{Downstream Application} & \multicolumn{1}{c}{StackOverflow} & Total \\ \midrule
2023.06.14                    & WizardCoder-15B-V1.0~\citep{abs-2306-08568}                                 & 56.7      & 50.6       & 64.3   & 54.2   & \multicolumn{1}{c}{0.17}                & \multicolumn{1}{c}{0}                      & \multicolumn{1}{c}{0.1}           & 0.06  \\ \midrule
2023.06.14                    & WizardCoder-Python-7B-V1.0~\citep{abs-2306-08568}                           & 50.6      & 45.1       & 58.5   & 49.5   & \multicolumn{1}{c}{6.62}                & \multicolumn{1}{c}{0.17}                   & \multicolumn{1}{c}{5.45}          & 2.66  \\ \midrule
2023.07.18                    & Llama-2-7B~\citep{abs-2307-09288}                                           & 12.8      & -          & 20.8   & -      & \multicolumn{1}{c}{6.57}                & \multicolumn{1}{c}{0.46}                   & \multicolumn{1}{c}{4.76}          & 2.74  \\ \midrule
2023.07.18                    & Llama-2-13B-Chat~\citep{abs-2307-09288}                                     & 18.3      & -          & 30.6   & -      & \multicolumn{1}{c}{3.71}                & \multicolumn{1}{c}{0.06}                   & \multicolumn{1}{c}{3.41}          & 1.51  \\ \midrule
2023.08.25                    & CodeLlama-7B-Instruct~\citep{abs-2308-12950}                                & 34.8      & -          & 44.4   & -      & \multicolumn{1}{c}{17.77}               & \multicolumn{1}{c}{0.62}                   & \multicolumn{1}{c}{17.8}          & 7.62  \\ \midrule
2023.08.25                    & CodeLlama-13B-Instruct~\citep{abs-2308-12950}                               & 42.7      & -          & 49.4   & -      & \multicolumn{1}{c}{28.45}               & \multicolumn{1}{c}{2.47}                   & \multicolumn{1}{c}{32.05}         & 13.5  \\ \midrule
2023.08.28                    & CodeLlama-7B-Python~\citep{abs-2308-12950}                                  & 38.4      & -          & 47.6   & -      & \multicolumn{1}{c}{3.4}                 & \multicolumn{1}{c}{0.03}                   & \multicolumn{1}{c}{2.35}          & 1.28  \\ \midrule
2023.10.29                    & DeepSeek-Coder-6.7B-Instruct~\citep{abs-2401-14196}                         & 74.4      & 71.3       & 74.9   & 65.6   & \multicolumn{1}{c}{3.83}                & \multicolumn{1}{c}{0.15}                   & \multicolumn{1}{c}{4.34}          & 1.71  \\ \midrule
2023.11.11                    & Mistral-7B-Instruct-V0.2~\citep{abs-2310-06825}                             & 42.1      & 36         & 44.7   & 37     & \multicolumn{1}{c}{13.96}               & \multicolumn{1}{c}{1.85}                   & \multicolumn{1}{c}{20.33}         & 7.54  \\ \midrule
2024.01.25                    & DeepSeek-Coder-7B-Instruct-V1.5~\citep{abs-2401-14196}                      & 75.6      & 71.3       & 75.2   & 62.2   & \multicolumn{1}{c}{26.7}                & \multicolumn{1}{c}{4.51}                   & \multicolumn{1}{c}{44.77}         & 15.71 \\ \midrule
2024.01.25                    & GPT-3.5-Turbo~\citep{gpt35}                               & 76.8      & 70.7       & 82.5   & 69.7   & \multicolumn{1}{c}{40.55}               & \multicolumn{1}{c}{30.48}                  & \multicolumn{1}{c}{65.95}         & 37.59 \\ \midrule
2024.02.27                    & StarCoder2-7B~\citep{abs-2402-19173}                                        & 35.4      & 29.9       & 55.4   & 45.6   & \multicolumn{1}{c}{12.21}               & \multicolumn{1}{c}{0.32}                   & \multicolumn{1}{c}{13.02}         & 5.27  \\ \midrule
2024.02.27                    & StarCoder2-15B~\citep{abs-2402-19173}                                       & 46.3      & 37.8       & 66.2   & 53.1   & \multicolumn{1}{c}{29.7}                & \multicolumn{1}{c}{2.9}                    & \multicolumn{1}{c}{35.79}         & 14.55 \\ \midrule
2024.04.09                    & CodeGemma-7B-Instruct~\citep{codegemma_2024}                                & 60.4      & 51.8       & 70.4   & 56.9   & \multicolumn{1}{c}{31.8}                & \multicolumn{1}{c}{0.76}                   & \multicolumn{1}{c}{31.29}         & 13.36 \\ \midrule
2024.04.09                    & CodeGemma-7B~\citep{codegemma_2024}                                         & 44.5      & 41.5       & 65.1   & 52.4   & \multicolumn{1}{c}{29.61}               & \multicolumn{1}{c}{1.12}                   & \multicolumn{1}{c}{34.01}         & 13.28 \\ \midrule
2024.04.10                    & aiXCoder-7B~\citep{aixcoder}      & 54.9      & -          & 66     & -      & \multicolumn{1}{c}{17.51}               & \multicolumn{1}{c}{1.09}                   & \multicolumn{1}{c}{26.3}          & 8.83  \\ \midrule
2024.04.15                    & aiXCoder-7B-Base~\citep{aixcoder} & 43.2      & -          & 62.2   & -      & \multicolumn{1}{c}{20.41}               & \multicolumn{1}{c}{0.94}                   & \multicolumn{1}{c}{26.37}         & 9.59  \\ \midrule
2024.04.15                    & CodeQwen1.5-7B~\citep{qwen}                                                 & 51.8      & 45.7       & 73.5   & 60.8   & \multicolumn{1}{c}{11.61}               & \multicolumn{1}{c}{0.12}                   & \multicolumn{1}{c}{7.58}          & 4.33  \\ \midrule
2024.04.15                    & CodeQwen1.5-7B-Chat~\citep{qwen}                                            & 83.5      & 78.7       & 79.4   & 69     & \multicolumn{1}{c}{12.16}               & \multicolumn{1}{c}{0.33}                   & \multicolumn{1}{c}{9.2}           & 4.81  \\ \midrule
2024.04.18                    & Llama-3-8B~\citep{llama3modelcard}                                          & 35.5      & 29.3       & 61.4   & 51.6   & \multicolumn{1}{c}{17.18}               & \multicolumn{1}{c}{0.24}                   & \multicolumn{1}{c}{20.69}         & 7.57  \\ \midrule
2024.04.18                    & Llama-3-8B-Instruct~\citep{llama3modelcard}                                 & 61.6      & 56.7       & 70.1   & 59.3   & \multicolumn{1}{c}{20.79}               & \multicolumn{1}{c}{3.67}                   & \multicolumn{1}{c}{34.08}         & 12.23 \\ \midrule
2024.04.18                    & Llama-3-70B-Chat~\citep{llama3modelcard}                                    & 77.4      & 72         & 82.3   & 69     & \multicolumn{1}{c}{33.76}               & \multicolumn{1}{c}{50.93}                  & \multicolumn{1}{c}{64.35}         & 47.55 \\ \midrule
2024.05.13                    & GPT-4o~\citep{gpt4o}                                      & 85.4      & 81.7       & 85.7   & 73.3   & \multicolumn{1}{c}{58.37}               & \multicolumn{1}{c}{72.98}                  & \multicolumn{1}{c}{87.21}         & 70.44 \\ \bottomrule
\end{tabular}
}
\caption{{\small Full evaluation results of EM@1 on token-level code completion compared to related datasets and different data sources. The results for related datasets are collected from the online leaderboard of Evalplus~\citep{LiuXW023}.\label{tab:main}}}
\end{table}

%% file: table/multi_language.tex
\begin{table}[t]
\centering
\resizebox{.9\textwidth}{!}{
\begin{tabular}{@{}ccccccccc@{}}
\toprule
                            \multirow{2}{*}{\textbf{Model}}                          & \multicolumn{2}{c}{\textbf{Python}}                     & \multicolumn{2}{c}{\textbf{Java}}                       & \multicolumn{2}{c}{\textbf{C\#}}                         & \multicolumn{2}{c}{\textbf{JavaScript}}                 \\ 
                                                      \cmidrule(lr){2-3}\cmidrule(lr){4-5}
                                                      \cmidrule(lr){6-7}
                                                      \cmidrule(lr){8-9}
                                                      & \multicolumn{1}{c}{ISM@1} & PM@1 & \multicolumn{1}{c}{ISM@1} & PM@1 & \multicolumn{1}{c}{ISM@1} & PM@1 & \multicolumn{1}{c}{ISM@1} & PM@1 \\ \midrule
DeepSeek-Coder-7B-Instruct-V1.5~\citep{abs-2401-14196} & \multicolumn{1}{c}{40.03}        & 27.35       & \multicolumn{1}{c}{61.55}        & 46.62       & \multicolumn{1}{c}{71.43}        & 49.68       & \multicolumn{1}{c}{75.22}        & 54.24       \\ \midrule
CodeLlama-13B-Instruct~\citep{abs-2308-12950}          & \multicolumn{1}{c}{48.83}        & 34.63       & \multicolumn{1}{c}{70.92}        & 58.87       & \multicolumn{1}{c}{47.62}        & 35.54       & \multicolumn{1}{c}{52.87}        & 34.11       \\ \midrule
StarCoder2-15B~\citep{abs-2402-19173}                  & \multicolumn{1}{c}{39.71}        & 27.36       & \multicolumn{1}{c}{38.63}        & 27.43       & \multicolumn{1}{c}{33.33}        & 28.63       & \multicolumn{1}{c}{60.67}        & 39.33       \\ \midrule
CodeGemma-7B~\citep{codegemma_2024}                    & \multicolumn{1}{c}{8.67}         & 5.00           & \multicolumn{1}{c}{34.38}        & 23.53       & \multicolumn{1}{c}{0}            & 0           & \multicolumn{1}{c}{16.82}        & 10.53       \\ \midrule
GPT-3.5-Turbo~\citep{gpt35}          & \multicolumn{1}{c}{40.77}        & 28.06       & \multicolumn{1}{c}{50.00}           & 39.34       & \multicolumn{1}{c}{28.57}        & 26.87       & \multicolumn{1}{c}{24.39}        & 15.85       \\ \midrule
GPT-4o~\citep{gpt4o}                 & \multicolumn{1}{c}{64.72}        & 50.48       & \multicolumn{1}{c}{70.83}        & 64.04       & \multicolumn{1}{c}{71.43}        & 63.26       & \multicolumn{1}{c}{77.74}        & 70.24       \\ \midrule
Llama-3-70B-Chat~\citep{llama3modelcard}               & \multicolumn{1}{c}{57.68}        & 41.47       & \multicolumn{1}{c}{61.55}        & 58.57       & \multicolumn{1}{c}{66.67}        & 56.35       & \multicolumn{1}{c}{75.61}        & 67.61       \\ \bottomrule
\end{tabular}
}
\caption{{\small Multi-language performance on VersiCode\label{tab:language}}}
\end{table}

%% file: table/wo_grammar_verfication.tex
\begin{table}[t]
\centering
\includegraphics[width=0.8\linewidth]{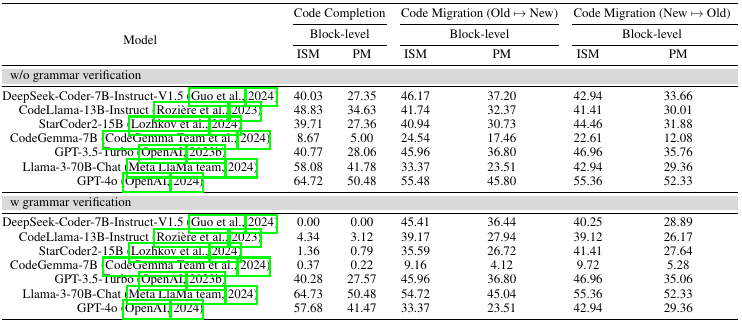}
\caption{{\small Results of block-level code completion and migration with or without grammar verification. \label{tab:wogrammar}}}
\end{table}

%% file: sections/D_metric_design.tex
\section{Metric Design of Critical Diff Check}
\label{apd:metric_design}

\subsection{Introduction of Critical Diff Check}
Critical Diff Check (CDC) focuses on the changes in the code rather than the overall similarity of the entire code segment. CDC has five rules as follows:

\begin{itemize}
    \item \emph{Rule 1: Check whether the generated code contains the core token.}
    \item \emph{Rule 2: Check whether the generated code is valid.} 
    \item \emph{Rule 3: Check if the number of arguments in the function using the core token is consistent.}
    \item \emph{Rule 4: If the reference code
uses a with statement, checks whether the generated code also uses a with statement.}
    \item \emph{Rule 5: If the reference
code uses keyword argument assignment, checks whether the generated code uses the same keyword argument
assignment.}
    
\end{itemize}

The failure frequency and examples for each rule are shown in Table~\ref{tab:metric_effectiveness}.


\input{table/metric_rule}

\subsection{Ablation Study of Critical Diff Check}
We conducted ablation experiments on the five CDC rules and calculated the Pearson correlation coefficient with the Pass@1 metric for each, to demonstrate the reliability of CDC. The specific experimental data is shown in Table~\ref{tab:metric_ablation}.

\label{subapd:ablation_metric}

\input{table/CDC_ablation_study}

%% file: table/metric_rule.tex
\begin{table*}[t]
\centering
\includegraphics[width=\linewidth]{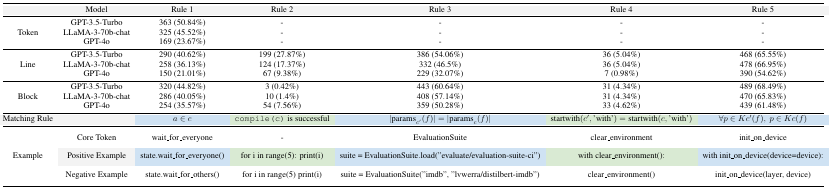}
\caption{{\small Each rule of the CDC, along with the frequency, occurrence rate, and examples of mismatches for each rule. `a' represents the core token, `c' represents the code generated by the model, `c$'$' represents the reference code, `f' represents the function of the specified token, and `params' refers to the function's parameter list. `Kc$'$(f)' and `Kc(f)' represent the keyword parameter lists of the reference code and the model-generated code, respectively, and `p' represents the parameter assigned using keyword arguments. In detail, \emph{Rule 1} checks whether the generated code contains the core token;
\emph{Rule 2} checks whether the generated code is valid; \emph{Rule 3} checks if the number of arguments in the function using the core token is consistent; \emph{Rule 4}, if the reference code uses a with statement, checks whether the generated code also uses a with statement; \emph{Rule 5}, if the reference code uses keyword argument assignment, checks whether the generated code uses the same keyword argument assignment. } \label{tab:metric_effectiveness}}
\end{table*}

%% file: table/CDC_ablation_study.tex
\begin{table*}[t]
\centering
\includegraphics[width=0.9\linewidth]{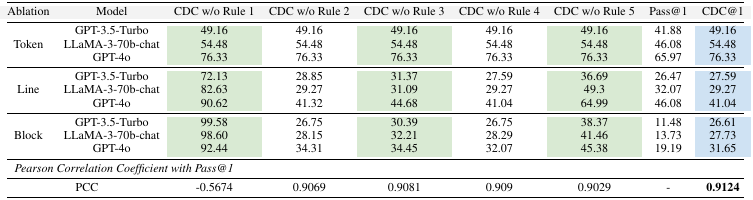}
\caption{{\small Ablation study of Critical Diff Check per rule. The configuration labeled as ``CDC w/o Rule i'', where $i \in\{1, 2, 3, 4, 5\}$ means that when calculating the CDC score, Rule i is excluded, and only the other four rules are considered. The Pearson correlation coefficient calculates the correlation the metric’s results obtained in each configuration against Pass@1.} \label{tab:metric_ablation}}
\end{table*}

%% file: sections/E_instance_example.tex
\section{Running Example of Executable Test}
\label{apd:example_execute}
\begin{figure*}
    \centering
    \includegraphics[width=.7\linewidth]{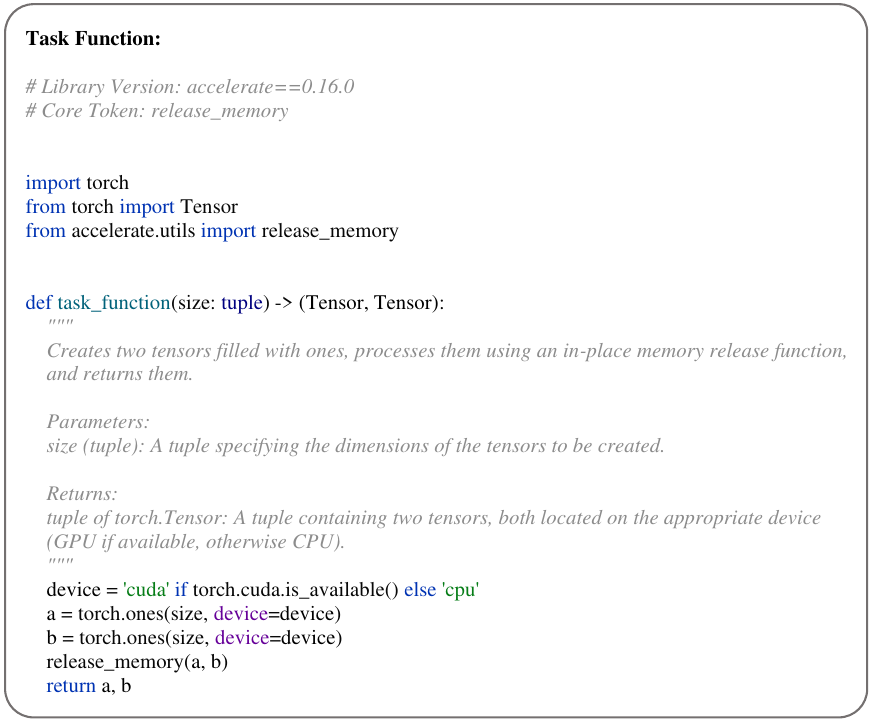}
    \caption{\small The ground truth for block-level code generation, used for Section~\ref{sect:line_level_completion}. Note that, ``core token'' is only provided for visualization, which is unseen for models. ``library version'' is optional, identified as ``w/ or w/o version'', and ``import'' statements are also optional, identified as ``w/ or w/o import'' in Table~\ref{tab:metric_gran_completion}.
    \label{fig:example_task_function}}
\end{figure*}

\begin{figure*}
    \centering
    \includegraphics[width=.7\linewidth]{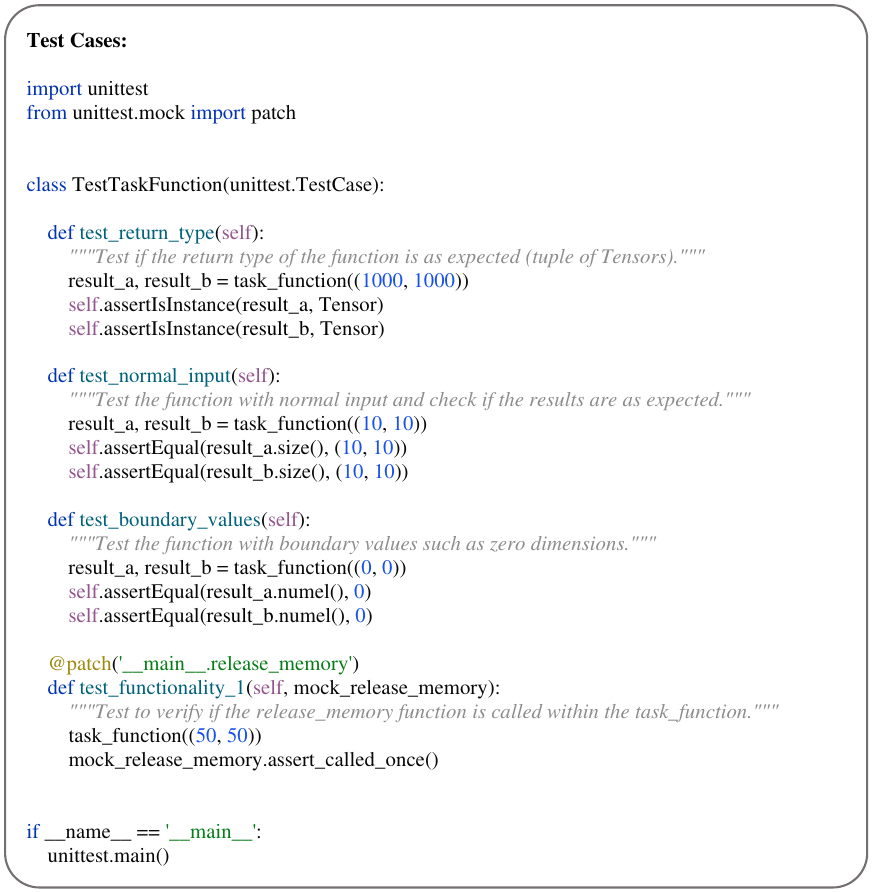}
    \caption{\small The test cases associated with generated code for dynamic code analysis, used for Section~\ref{sect:line_level_completion}.
    \label{fig:example_test_cases}}
\end{figure*}

As shown in Figure~\ref{fig:example_task_function},  this is an example of a task function used for code generation, where the task function is processed in various granular forms of code completion. The ``core token'' is only provided for visualization, which is unseen for models. ``library version'' is optional, identified as ``w/ or w/o version'', and ``import'' statements are also optional, identified as ``w/ or w/o import'' in Table~\ref{tab:metric_gran_completion}. As shown in Figure~\ref{fig:example_test_cases}, these are the test cases for the task function illustrated in Figure~\ref{fig:example_task_function}. The test cases were developed by experts through interactions with GPT-4 and include four types of tests. 

%% file: sections/F_prompt_template.tex
\section{Evaluation Details}
\label{apd:evaluation_details}
\subsection{Hyper-parameter}
\label{subapd:parameter}

As illustrated in Table~\ref{tab:hyper}, we have itemized the hyper-parameters pertinent to version-controllable code generation.

\begin{table}[t]
\centering
\resizebox{.6\textwidth}{!}{
\begin{tabular}{@{}ccccc@{}}
\toprule
\multirow{2}{*}{\textbf{hyper-parameter}} & \multicolumn{3}{c}{\textbf{code completion}} & \textbf{code migration} \\
\cmidrule(l){2-4}\cmidrule(l){5-5}
                                 & token-level   & line-level   & block-level   & block-level           \\
                            \midrule
temperature                      & 0.8           & 0.8          & 0.8           & 0.8                   \\
top\_p                           & 0.95          & 0.95         & 0.95          & 0.95                  \\
max\_tokens                      & 64            & 128          & 512           & 512                   \\
n                                & 100           & 6            & 6             & 6     \\      
\bottomrule
\end{tabular}}
\caption{{\small Hyper-parameters for completion and migration.\label{tab:hyper}}}
\end{table}

\subsection{Prompt Template}
\label{subapd:prompt}

\begin{figure*}[t]
    \centering
    \includegraphics[width=.8\textwidth]{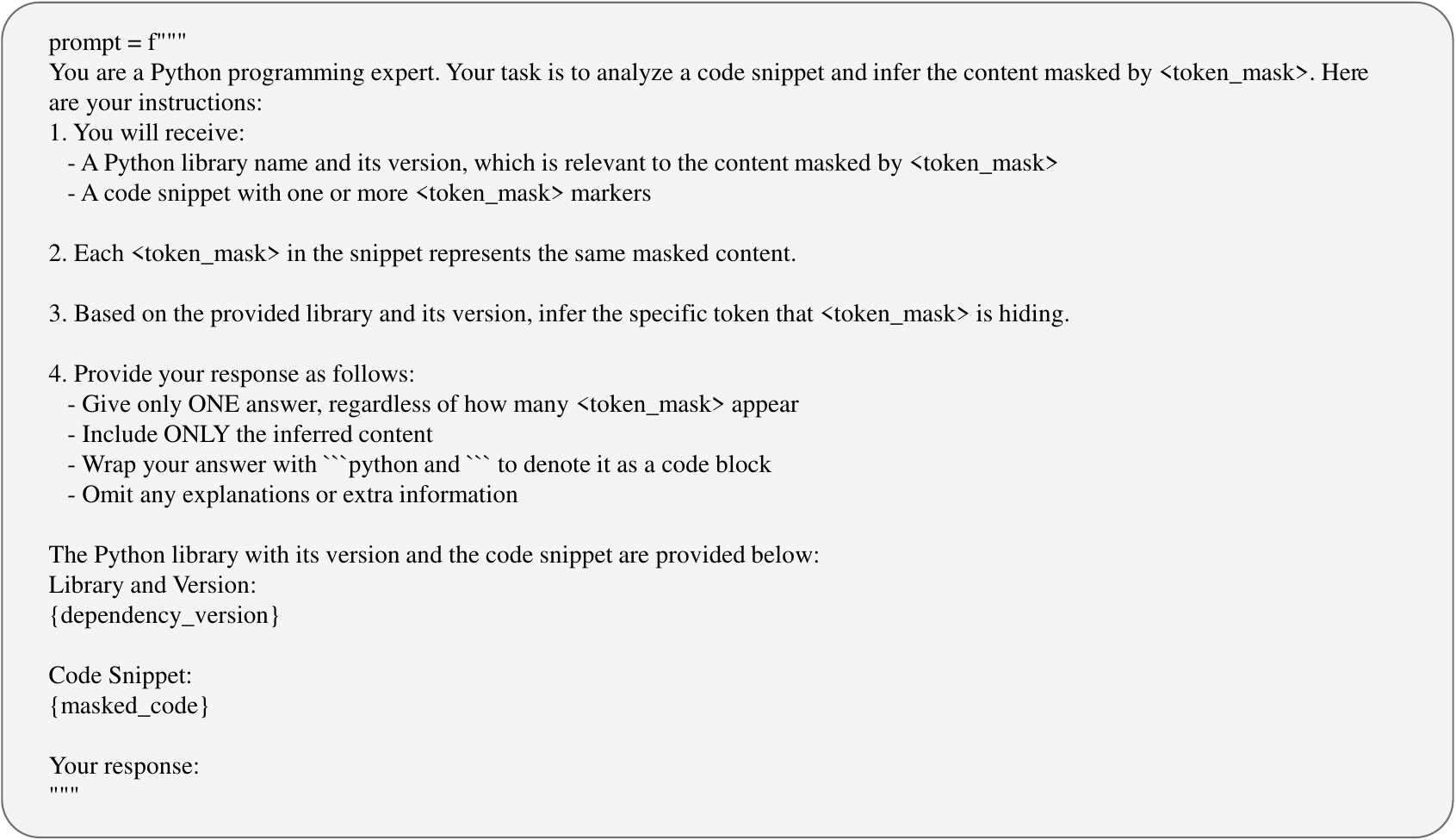}
    \caption{{\small Prompt template for token-level version-specific code completion.}}
    \label{fig:token_prompt}
\end{figure*}

\begin{figure*}[t]
    \centering
    \includegraphics[width=.8\textwidth]{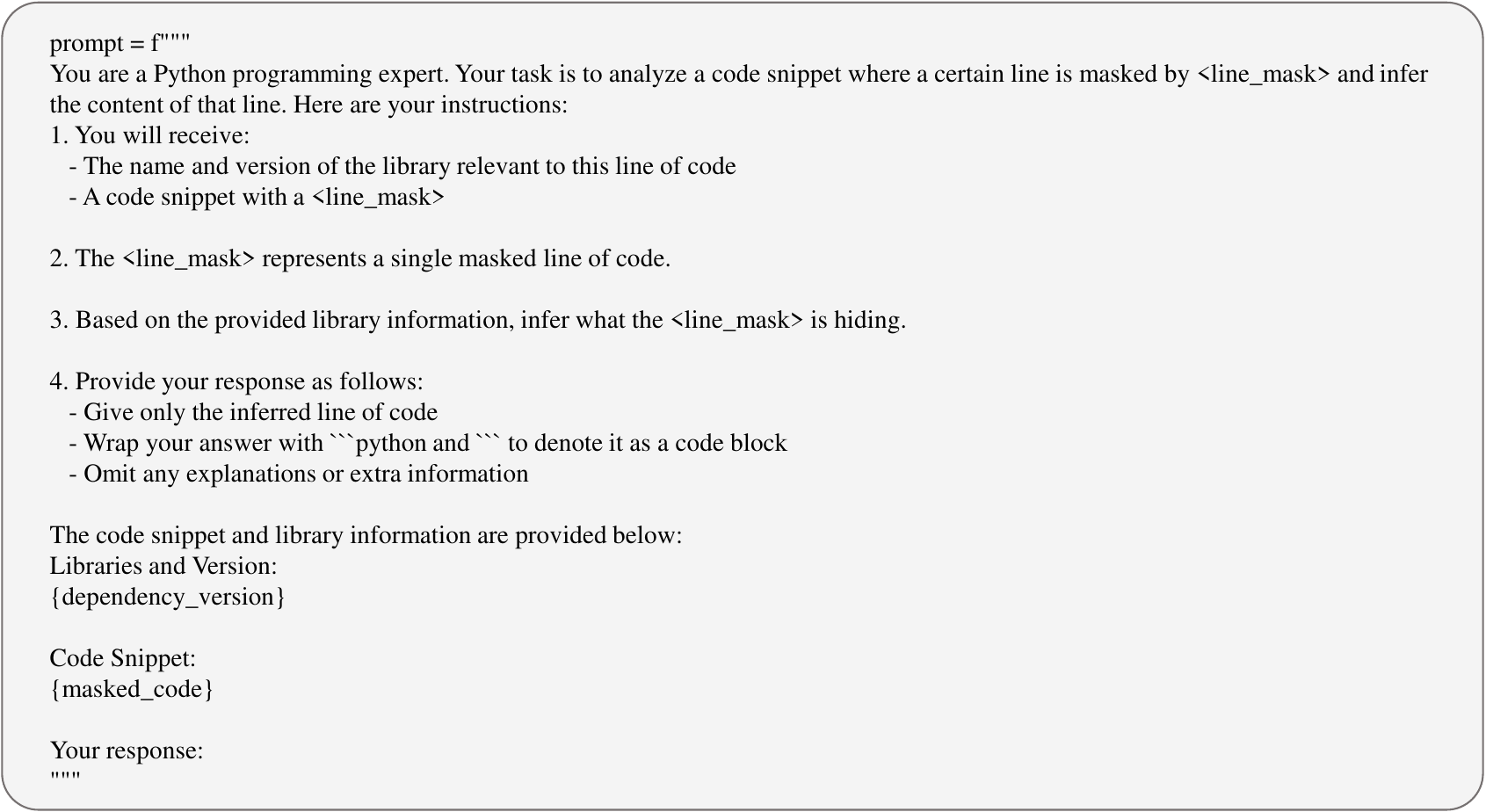}
    \caption{{\small Prompt template for line-level version-specific code completion.}}
    \label{fig:line-template}
\end{figure*}

\begin{figure*}[t]
    \centering
    \includegraphics[width=.8\textwidth]{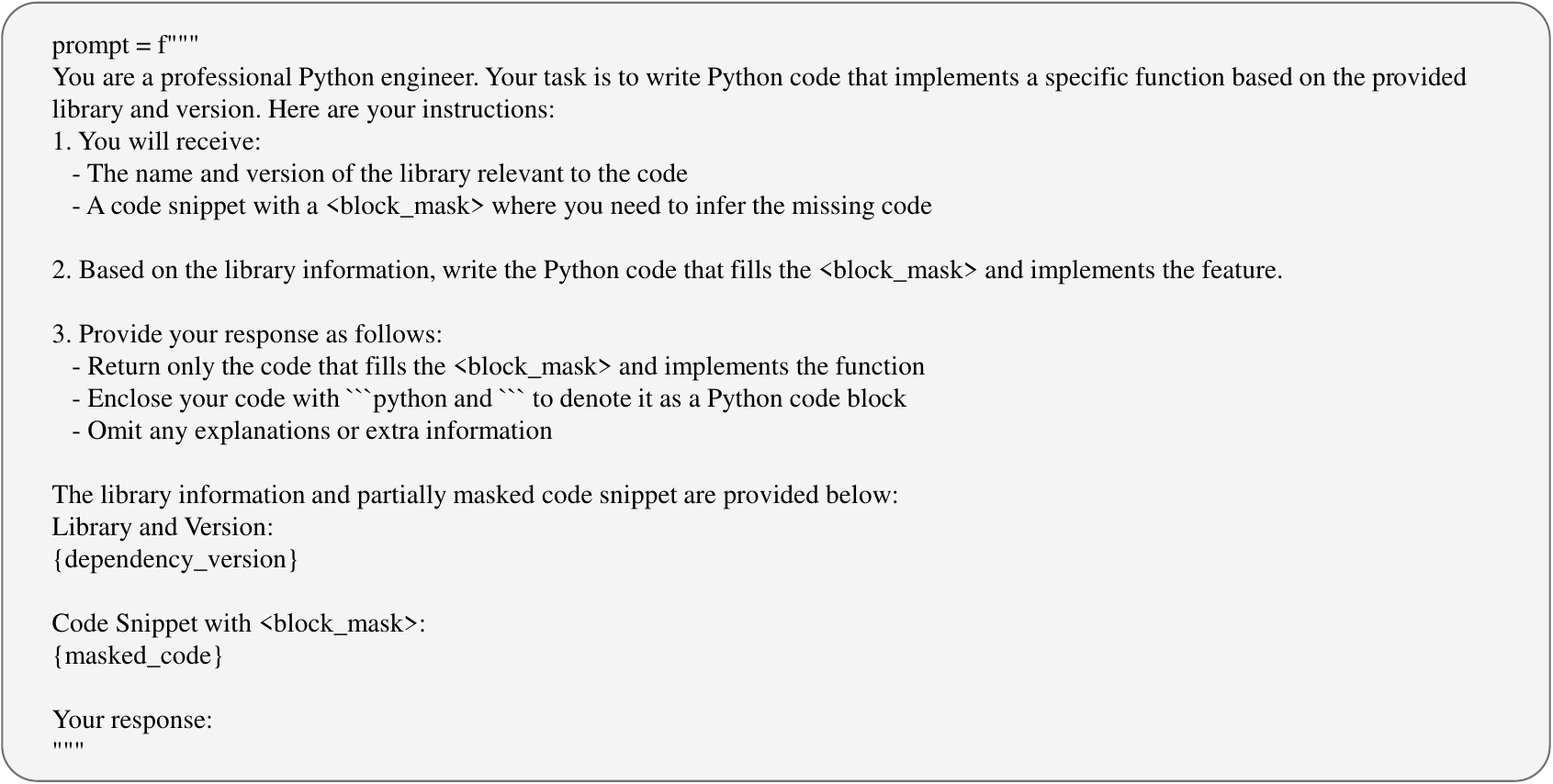}
    \caption{{\small Prompt template for block-level version-specific code completion.}}
    \label{fig:block-template}
\end{figure*}

We introduce the prompt template for token-level, line-level, and block-level evaluations in Figure~\ref{fig:token_prompt}, Figure~\ref{fig:line-template}, and Figure~\ref{fig:block-template}, respectively.

\begin{figure*}
    \centering
    \includegraphics[width=.8\textwidth]{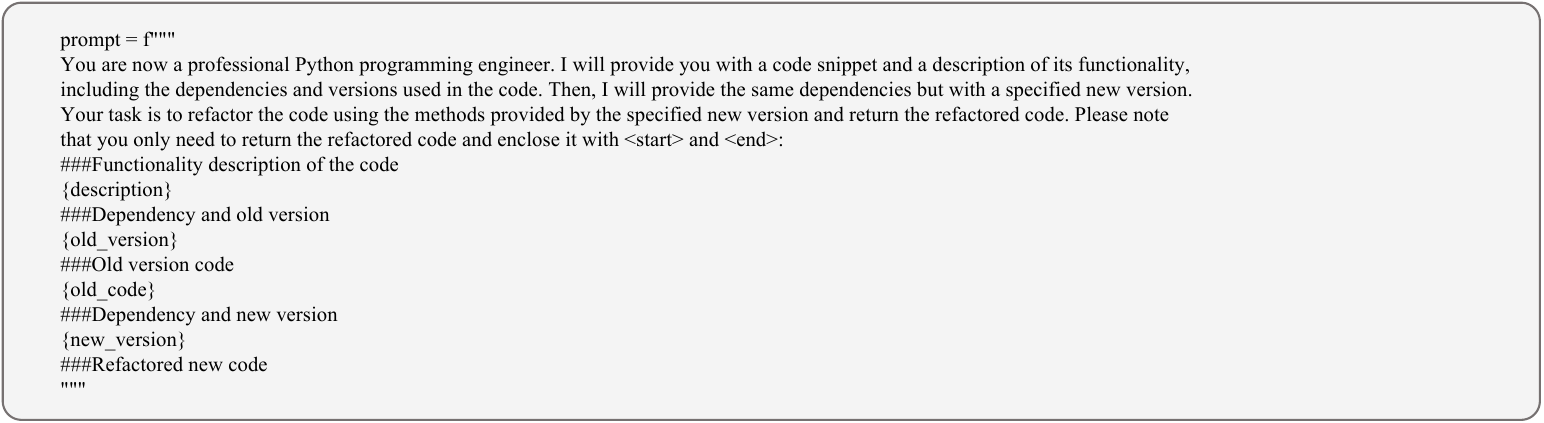}
    \caption{{\small Prompt template for version-aware code migration.}}
    \label{fig:enter-label}
\end{figure*}

\subsection{Data Sampling}
\label{subapd:sample}
For token-level completion tasks(Figure~\ref{fig:main}), we randomly sampled 2,000 instances for evaluation. We used the entire executable dataset for line- and block-level completion tasks due to its smaller size (Figure~\ref{fig:test_cases_performance}, Table~\ref{tab:metric_gran_completion}). In the time trend experiment (Figure~\ref{fig:time_analysis}), we sampled 200 data points per quarter or used all available data if fewer. And in the code migration task (Table~\ref{tab:migration_results}), we randomly sampled 2,000 instances for evaluation.

%% file: sections/G_error_analysis.tex
\section{Error Analysis}
\label{apd:negative_example}
\subsection{Error Analysis of GPT4-o}
\label{subapd:ea}
\begin{figure}[t]
    \centering
    \includegraphics[width=.8\textwidth, bb=0 0 741.47 273.6]{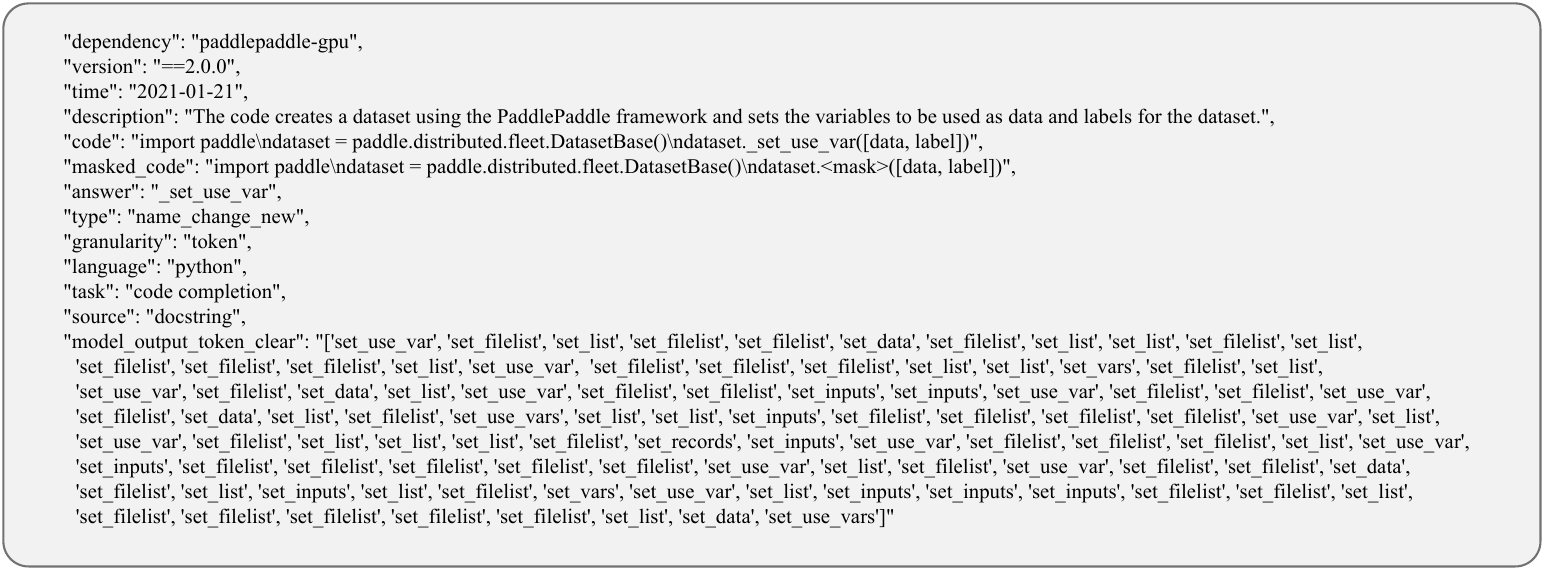}
    \caption{{\small The first negative example of GPT-4o on token-level code completion.}}
    \label{fig:ne1}
\end{figure}

\begin{figure}[t]
    \centering
    \includegraphics[width=.8\textwidth, bb=0 0 741.47 225.6]{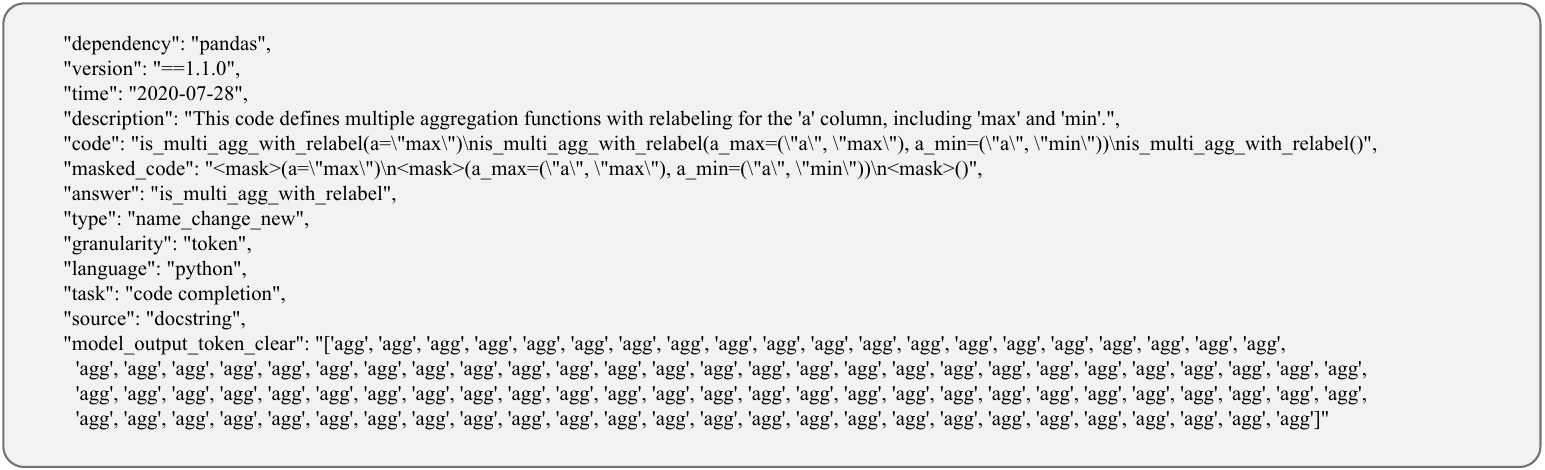}
    \caption{{\small The second negative example of GPT-4o on token-level code completion.}}
    \label{fig:ne2}
\end{figure}

\begin{figure}[t]
    \centering
    \includegraphics[width=.8\textwidth, bb=0 0 743.877 296.4]{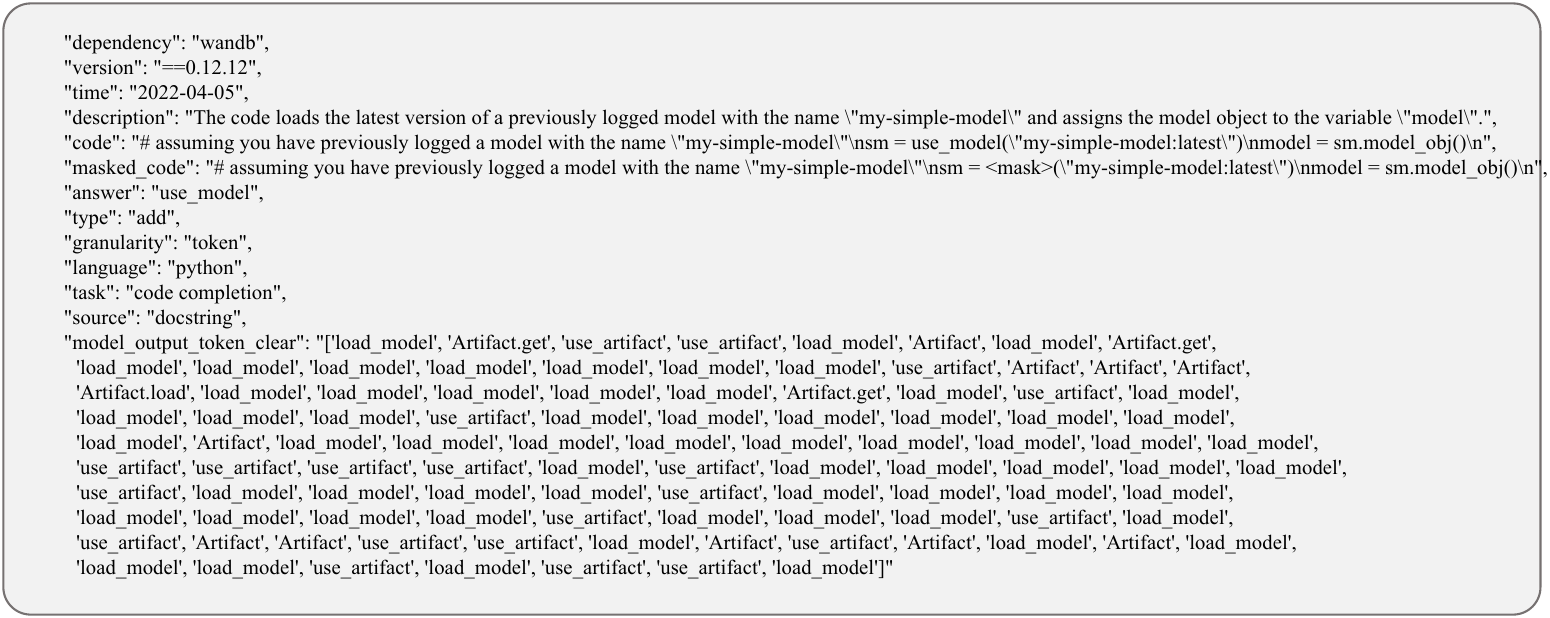}
    \caption{{\small The third negative example of GPT-4o on token-level code completion.}}
    \label{fig:ne3}
\end{figure}

Despite GPT4-o achieving superior performance in general evaluation, it still encounters errors in 30\% of instances. We provide several negative examples in Figure~\ref{fig:ne1}, Figure~\ref{fig:ne2}, and Figure~\ref{fig:ne3}.

%% file: iclr2025_conference.bbl
\begin{thebibliography}{64}
\providecommand{\natexlab}[1]{#1}
\providecommand{\url}[1]{\texttt{#1}}
\expandafter\ifx\csname urlstyle\endcsname\relax
  \providecommand{\doi}[1]{doi: #1}\else
  \providecommand{\doi}{doi: \begingroup \urlstyle{rm}\Url}\fi

\bibitem[Agrawal et~al.(2023)Agrawal, Kanade, Goyal, Lahiri, and Rajamani]{abs-2306-10763}
Lakshya~A Agrawal, Aditya Kanade, Navin Goyal, Shuvendu~K. Lahiri, and Sriram~K. Rajamani.
\newblock Guiding language models of code with global context using monitors.
\newblock \emph{CoRR}, abs/2306.10763, 2023.
\newblock URL \url{https://doi.org/10.48550/arXiv.2306.10763}.

\bibitem[Ahmad et~al.(2023)Ahmad, Tushar, Chakraborty, and Chang]{AhmadTCC23}
Wasi~Uddin Ahmad, Md~Golam~Rahman Tushar, Saikat Chakraborty, and Kai{-}Wei Chang.
\newblock {AVATAR:} {A} parallel corpus for java-python program translation.
\newblock In \emph{Findings of ACL}, pp.\  2268--2281, 2023.
\newblock URL \url{https://doi.org/10.18653/v1/2023.findings-acl.143}.

\bibitem[{aiXcoder team}(2024)]{aixcoder}
{aiXcoder team}.
\newblock aixcoder-7b code large language model.
\newblock \url{https://github.com/aixcoder-plugin/aiXcoder-7B}, 2024.
\newblock Accessed: June 7, 2024.

\bibitem[Austin et~al.(2021)Austin, Odena, Nye, Bosma, Michalewski, Dohan, Jiang, Cai, Terry, Le, and Sutton]{abs-2108-07732}
Jacob Austin, Augustus Odena, Maxwell~I. Nye, Maarten Bosma, Henryk Michalewski, David Dohan, Ellen Jiang, Carrie~J. Cai, Michael Terry, Quoc~V. Le, and Charles Sutton.
\newblock Program synthesis with large language models.
\newblock \emph{CoRR}, abs/2108.07732, 2021.
\newblock URL \url{https://arxiv.org/abs/2108.07732}.

\bibitem[Bai et~al.(2023)Bai, Bai, Chu, Cui, Dang, Deng, Fan, Ge, Han, Huang, et~al.]{qwen}
Jinze Bai, Shuai Bai, Yunfei Chu, Zeyu Cui, Kai Dang, Xiaodong Deng, Yang Fan, Wenbin Ge, Yu~Han, Fei Huang, et~al.
\newblock Qwen technical report.
\newblock \emph{CoRR}, abs/2309.16609, 2023.
\newblock URL \url{https://doi.org/10.48550/arXiv.2309.16609}.

\bibitem[Brito et~al.(2018)Brito, Hora, Valente, and Robbes]{BritoHVR18}
Gleison Brito, Andr{\'{e}}~C. Hora, Marco~T{\'{u}}lio Valente, and Romain Robbes.
\newblock On the use of replacement messages in {API} deprecation: An empirical study.
\newblock \emph{J. Syst. Softw.}, 137:\penalty0 306--321, 2018.
\newblock URL \url{https://doi.org/10.1016/j.jss.2017.12.007}.

\bibitem[Chen et~al.(2021)Chen, Tworek, Jun, Yuan, de~Oliveira~Pinto, Kaplan, Edwards, Burda, Joseph, Brockman, et~al.]{abs-2107-03374}
Mark Chen, Jerry Tworek, Heewoo Jun, Qiming Yuan, Henrique~Pond{\'{e}} de~Oliveira~Pinto, Jared Kaplan, Harrison Edwards, Yuri Burda, Nicholas Joseph, Greg Brockman, et~al.
\newblock Evaluating large language models trained on code.
\newblock \emph{CoRR}, abs/2107.03374, 2021.
\newblock URL \url{https://arxiv.org/abs/2107.03374}.

\bibitem[{CodeGemma Team} et~al.(2024){CodeGemma Team}, Hartman, Hu, Choquette-Choo, Zhao, Fine, Hui, et~al.]{codegemma_2024}
{CodeGemma Team}, Ale~Jakse Hartman, Andrea Hu, Christopher~A. Choquette-Choo, Heri Zhao, Jane Fine, Jeffrey Hui, et~al.
\newblock Codegemma: Open code models based on gemma.
\newblock \emph{Google}, 2024.
\newblock URL \url{https://goo.gle/codegemma}.

\bibitem[Dig \& Johnson(2006)Dig and Johnson]{DigJ06}
Danny Dig and Ralph~E. Johnson.
\newblock How do apis evolve? {A} story of refactoring.
\newblock \emph{J. Softw. Maintenance Res. Pract.}, 18\penalty0 (2):\penalty0 83--107, 2006.
\newblock URL \url{https://doi.org/10.1002/smr.328}.

\bibitem[Dilhara et~al.(2021)Dilhara, Ketkar, and Dig]{DilharaKD21}
Malinda Dilhara, Ameya Ketkar, and Danny Dig.
\newblock Understanding software-2.0: {A} study of machine learning library usage and evolution.
\newblock \emph{{ACM} Trans. Softw. Eng. Methodol.}, 30\penalty0 (4):\penalty0 55:1--55:42, 2021.
\newblock URL \url{https://doi.org/10.1145/3453478}.

\bibitem[Du et~al.(2024)Du, Rui, Chai, Fu, Xia, Wang, Tang, Yu, and Zhang]{abs-2405-02355}
Kounianhua Du, Renting Rui, Huacan Chai, Lingyue Fu, Wei Xia, Yasheng Wang, Ruiming Tang, Yong Yu, and Weinan Zhang.
\newblock Codegrag: Extracting composed syntax graphs for retrieval augmented cross-lingual code generation.
\newblock \emph{CoRR}, abs/2405.02355, 2024.
\newblock URL \url{https://doi.org/10.48550/arXiv.2405.02355}.

\bibitem[Feng et~al.(2020)Feng, Guo, Tang, Duan, Feng, Gong, Shou, Qin, Liu, Jiang, and Zhou]{FengGTDFGS0LJZ20}
Zhangyin Feng, Daya Guo, Duyu Tang, Nan Duan, Xiaocheng Feng, Ming Gong, Linjun Shou, Bing Qin, Ting Liu, Daxin Jiang, and Ming Zhou.
\newblock Codebert: {A} pre-trained model for programming and natural languages.
\newblock In \emph{Findings of EMNLP}, pp.\  1536--1547, 2020.
\newblock URL \url{https://doi.org/10.18653/v1/2020.findings-emnlp.139}.

\bibitem[Fried et~al.(2023)Fried, Aghajanyan, Lin, Wang, Wallace, Shi, Zhong, Yih, Zettlemoyer, and Lewis]{FriedAL0WSZYZL23}
Daniel Fried, Armen Aghajanyan, Jessy Lin, Sida Wang, Eric Wallace, Freda Shi, Ruiqi Zhong, Scott Yih, Luke Zettlemoyer, and Mike Lewis.
\newblock Incoder: {A} generative model for code infilling and synthesis.
\newblock In \emph{ICLR}, 2023.
\newblock URL \url{https://openreview.net/pdf?id=hQwb-lbM6EL}.

\bibitem[Fu et~al.(2023)Fu, Chai, Luo, Du, Zhang, Fan, Lei, Rui, Lin, Fang, et~al.]{abs-2309-01940}
Lingyue Fu, Huacan Chai, Shuang Luo, Kounianhua Du, Weiming Zhang, Longteng Fan, Jiayi Lei, Renting Rui, Jianghao Lin, Yuchen Fang, et~al.
\newblock Codeapex: {A} bilingual programming evaluation benchmark for large language models.
\newblock \emph{CoRR}, abs/2309.01940, 2023.
\newblock URL \url{https://doi.org/10.48550/arXiv.2309.01940}.

\bibitem[Gao et~al.(2023)Gao, Xiong, Gao, Jia, Pan, Bi, Dai, Sun, Guo, Wang, and Wang]{abs-2312-10997}
Yunfan Gao, Yun Xiong, Xinyu Gao, Kangxiang Jia, Jinliu Pan, Yuxi Bi, Yi~Dai, Jiawei Sun, Qianyu Guo, Meng Wang, and Haofen Wang.
\newblock Retrieval-augmented generation for large language models: {A} survey.
\newblock \emph{CoRR}, abs/2312.10997, 2023.
\newblock URL \url{https://doi.org/10.48550/arXiv.2312.10997}.

\bibitem[Gunasekar et~al.(2023)Gunasekar, Zhang, Aneja, Mendes, Giorno, Gopi, Javaheripi, Kauffmann, de~Rosa, Saarikivi, et~al.]{abs-2306-11644}
Suriya Gunasekar, Yi~Zhang, Jyoti Aneja, Caio C{\'{e}}sar~Teodoro Mendes, Allie~Del Giorno, Sivakanth Gopi, Mojan Javaheripi, Piero Kauffmann, Gustavo de~Rosa, Olli Saarikivi, et~al.
\newblock Textbooks are all you need.
\newblock \emph{CoRR}, abs/2306.11644, 2023.
\newblock URL \url{https://doi.org/10.48550/arXiv.2306.11644}.

\bibitem[Guo et~al.(2024)Guo, Zhu, Yang, Xie, Dong, Zhang, Chen, Bi, Wu, Li, et~al.]{abs-2401-14196}
Daya Guo, Qihao Zhu, Dejian Yang, Zhenda Xie, Kai Dong, Wentao Zhang, Guanting Chen, Xiao Bi, Y.~Wu, Y.~K. Li, et~al.
\newblock Deepseek-coder: When the large language model meets programming - the rise of code intelligence.
\newblock \emph{CoRR}, abs/2401.14196, 2024.
\newblock URL \url{https://doi.org/10.48550/arXiv.2401.14196}.

\bibitem[Haryono et~al.(2021)Haryono, Thung, Lo, Lawall, and Jiang]{HaryonoT0LJ21}
Stefanus~A. Haryono, Ferdian Thung, David Lo, Julia Lawall, and Lingxiao Jiang.
\newblock Characterization and automatic updates of deprecated machine-learning {API} usages.
\newblock In \emph{Proceedings of ICSME}, pp.\  137--147. {IEEE}, 2021.
\newblock URL \url{https://doi.org/10.1109/ICSME52107.2021.00019}.

\bibitem[Hendrycks et~al.(2021)Hendrycks, Basart, Kadavath, Mazeika, Arora, Guo, Burns, Puranik, He, Song, and Steinhardt]{HendrycksBKMAGB21}
Dan Hendrycks, Steven Basart, Saurav Kadavath, Mantas Mazeika, Akul Arora, Ethan Guo, Collin Burns, Samir Puranik, Horace He, Dawn Song, and Jacob Steinhardt.
\newblock Measuring coding challenge competence with {APPS}.
\newblock In \emph{Proceedings of NeurIPS}, 2021.
\newblock URL \url{https://datasets-benchmarks-proceedings.neurips.cc/paper/2021/hash/c24cd76e1ce41366a4bbe8a49b02a028-Abstract-round2.html}.

\bibitem[Hu et~al.(2023)Hu, Li, Zhao, Xie, Liu, Chen, Xie, and He]{abs-2310-20329}
Qisheng Hu, Kaixin Li, Xu~Zhao, Yuxi Xie, Tiedong Liu, Hui Chen, Qizhe Xie, and Junxian He.
\newblock Instructcoder: Empowering language models for code editing.
\newblock \emph{CoRR}, abs/2310.20329, 2023.
\newblock URL \url{https://doi.org/10.48550/arXiv.2310.20329}.

\bibitem[Jang et~al.(2022)Jang, Ye, Lee, Yang, Shin, Han, Kim, and Seo]{JangYLYSHKS22}
Joel Jang, Seonghyeon Ye, Changho Lee, Sohee Yang, Joongbo Shin, Janghoon Han, Gyeonghun Kim, and Minjoon Seo.
\newblock Temporalwiki: {A} lifelong benchmark for training and evaluating ever-evolving language models.
\newblock In \emph{Proceedings of EMNLP}, pp.\  6237--6250, 2022.
\newblock URL \url{https://doi.org/10.18653/v1/2022.emnlp-main.418}.

\bibitem[Jiang et~al.(2023)Jiang, Sablayrolles, Mensch, Bamford, Chaplot, de~Las~Casas, Bressand, Lengyel, Lample, Saulnier, et~al.]{abs-2310-06825}
Albert~Q. Jiang, Alexandre Sablayrolles, Arthur Mensch, Chris Bamford, Devendra~Singh Chaplot, Diego de~Las~Casas, Florian Bressand, Gianna Lengyel, Guillaume Lample, Lucile Saulnier, et~al.
\newblock Mistral 7b.
\newblock \emph{CoRR}, abs/2310.06825, 2023.
\newblock URL \url{https://doi.org/10.48550/arXiv.2310.06825}.

\bibitem[Jiang et~al.(2024)Jiang, Wang, Shen, Kim, and Kim]{jiang2024survey}
Juyong Jiang, Fan Wang, Jiasi Shen, Sungju Kim, and Sunghun Kim.
\newblock A survey on large language models for code generation.
\newblock \emph{CoRR}, abs/2406.00515, 2024.
\newblock URL \url{https://doi.org/10.48550/arXiv.2406.00515}.

\bibitem[Jiao et~al.(2023)Jiao, Yu, Li, Qiu, Gu, and Shen]{JiaoYLQGS23}
Mingsheng Jiao, Tingrui Yu, Xuan Li, Guanjie Qiu, Xiaodong Gu, and Beijun Shen.
\newblock On the evaluation of neural code translation: Taxonomy and benchmark.
\newblock In \emph{Proceedings of ASE}, pp.\  1529--1541, 2023.
\newblock URL \url{https://doi.org/10.1109/ASE56229.2023.00114}.

\bibitem[Just et~al.(2014)Just, Jalali, and Ernst]{JustJE14}
Ren{\'{e}} Just, Darioush Jalali, and Michael~D. Ernst.
\newblock Defects4j: a database of existing faults to enable controlled testing studies for java programs.
\newblock In \emph{International Symposium on Software Testing and Analysis}, pp.\  437--440, 2014.
\newblock URL \url{https://doi.org/10.1145/2610384.2628055}.

\bibitem[Lai et~al.(2023)Lai, Li, Wang, Zhang, Zhong, Zettlemoyer, Yih, Fried, Wang, and Yu]{Lai0WZZZYFWY23}
Yuhang Lai, Chengxi Li, Yiming Wang, Tianyi Zhang, Ruiqi Zhong, Luke Zettlemoyer, Wen{-}Tau Yih, Daniel Fried, Sida~I. Wang, and Tao Yu.
\newblock {DS-1000:} {A} natural and reliable benchmark for data science code generation.
\newblock In \emph{Proceedings of ICML}, volume 202, pp.\  18319--18345, 2023.
\newblock URL \url{https://proceedings.mlr.press/v202/lai23b.html}.

\bibitem[Lamothe et~al.(2022)Lamothe, Gu{\'{e}}h{\'{e}}neuc, and Shang]{LamotheGS22}
Maxime Lamothe, Yann{-}Ga{\"{e}}l Gu{\'{e}}h{\'{e}}neuc, and Weiyi Shang.
\newblock A systematic review of {API} evolution literature.
\newblock \emph{{ACM} Comput. Surv.}, 54\penalty0 (8):\penalty0 171:1--171:36, 2022.
\newblock URL \url{https://doi.org/10.1145/3470133}.

\bibitem[Li et~al.(2023)Li, Bubeck, Eldan, Giorno, Gunasekar, and Lee]{abs-2309-05463}
Yuanzhi Li, S{\'{e}}bastien Bubeck, Ronen Eldan, Allie~Del Giorno, Suriya Gunasekar, and Yin~Tat Lee.
\newblock Textbooks are all you need {II:} phi-1.5 technical report.
\newblock \emph{CoRR}, abs/2309.05463, 2023.
\newblock URL \url{https://doi.org/10.48550/arXiv.2309.05463}.

\bibitem[Li et~al.(2022)Li, Choi, Chung, Kushman, Schrittwieser, Leblond, Eccles, Keeling, Gimeno, Lago, et~al.]{abs-2203-07814}
Yujia Li, David~H. Choi, Junyoung Chung, Nate Kushman, Julian Schrittwieser, R{\'{e}}mi Leblond, Tom Eccles, James Keeling, Felix Gimeno, Agustin~Dal Lago, et~al.
\newblock Competition-level code generation with alphacode.
\newblock \emph{CoRR}, abs/2203.07814, 2022.
\newblock URL \url{https://doi.org/10.48550/arXiv.2203.07814}.

\bibitem[Lin et~al.(2017)Lin, Koppel, Chen, and Solar{-}Lezama]{LinKCS17}
Derrick Lin, James Koppel, Angela Chen, and Armando Solar{-}Lezama.
\newblock Quixbugs: a multi-lingual program repair benchmark set based on the quixey challenge.
\newblock In \emph{Proceedings of SIGPLAN}, pp.\  55--56, 2017.
\newblock URL \url{https://doi.org/10.1145/3135932.3135941}.

\bibitem[Liu et~al.(2023)Liu, Xia, Wang, and Zhang]{LiuXW023}
Jiawei Liu, Chunqiu~Steven Xia, Yuyao Wang, and Lingming Zhang.
\newblock Is your code generated by chatgpt really correct? rigorous evaluation of large language models for code generation.
\newblock In \emph{Proceedings of NeurIPS}, 2023.
\newblock URL \url{http://papers.nips.cc/paper\_files/paper/2023/hash/43e9d647ccd3e4b7b5baab53f0368686-Abstract-Conference.html}.

\bibitem[Liu et~al.(2021)Liu, Li, Yan, Fazzini, and Grundy]{Liu0YFG21}
Pei Liu, Li~Li, Yichun Yan, Mattia Fazzini, and John~C. Grundy.
\newblock Identifying and characterizing silently-evolved methods in the android {API}.
\newblock In \emph{Proceedings of ICSE}, pp.\  308--317. {IEEE}, 2021.
\newblock URL \url{https://doi.org/10.1109/ICSE-SEIP52600.2021.00040}.

\bibitem[Lozhkov et~al.(2024)Lozhkov, Li, Allal, Cassano, Lamy{-}Poirier, Tazi, Tang, Pykhtar, Liu, Wei, Liu, Tian, et~al.]{abs-2402-19173}
Anton Lozhkov, Raymond Li, Loubna~Ben Allal, Federico Cassano, Joel Lamy{-}Poirier, Nouamane Tazi, Ao~Tang, Dmytro Pykhtar, Jiawei Liu, Yuxiang Wei, Tianyang Liu, Max Tian, et~al.
\newblock Starcoder 2 and the stack v2: The next generation.
\newblock \emph{CoRR}, abs/2402.19173, 2024.
\newblock URL \url{https://doi.org/10.48550/arXiv.2402.19173}.

\bibitem[Lu et~al.(2021)Lu, Guo, Ren, Huang, Svyatkovskiy, Blanco, Clement, Drain, Jiang, Tang, et~al.]{LuGRHSBCDJTLZSZ21}
Shuai Lu, Daya Guo, Shuo Ren, Junjie Huang, Alexey Svyatkovskiy, Ambrosio Blanco, Colin~B. Clement, Dawn Drain, Daxin Jiang, Duyu Tang, et~al.
\newblock Codexglue: {A} machine learning benchmark dataset for code understanding and generation.
\newblock In \emph{Proceedings of NeurIPS}, 2021.
\newblock URL \url{https://datasets-benchmarks-proceedings.neurips.cc/paper/2021/hash/c16a5320fa475530d9583c34fd356ef5-Abstract-round1.html}.

\bibitem[Luo et~al.(2024{\natexlab{a}})Luo, Ye, Liang, Zhang, Qin, Lu, Wu, Cong, Lin, Zhang, et~al.]{abs-2402-16667}
Qinyu Luo, Yining Ye, Shihao Liang, Zhong Zhang, Yujia Qin, Yaxi Lu, Yesai Wu, Xin Cong, Yankai Lin, Yingli Zhang, et~al.
\newblock Repoagent: An llm-powered open-source framework for repository-level code documentation generation.
\newblock \emph{CoRR}, abs/2402.16667, 2024{\natexlab{a}}.
\newblock URL \url{https://doi.org/10.48550/arXiv.2402.16667}.

\bibitem[Luo et~al.(2024{\natexlab{b}})Luo, Zhu, Zhang, Wang, Yang, Xu, and Che]{abs-2403-00338}
Xianzhen Luo, Qingfu Zhu, Zhiming Zhang, Xu~Wang, Qing Yang, Dongliang Xu, and Wanxiang Che.
\newblock Semi-instruct: Bridging natural-instruct and self-instruct for code large language models.
\newblock \emph{CoRR}, abs/2403.00338, 2024{\natexlab{b}}.
\newblock URL \url{https://doi.org/10.48550/arXiv.2403.00338}.

\bibitem[Luo et~al.(2024{\natexlab{c}})Luo, Xu, Zhao, Sun, Geng, Hu, Tao, Ma, Lin, and Jiang]{abs-2306-08568}
Ziyang Luo, Can Xu, Pu~Zhao, Qingfeng Sun, Xiubo Geng, Wenxiang Hu, Chongyang Tao, Jing Ma, Qingwei Lin, and Daxin Jiang.
\newblock Wizardcoder: Empowering code large language models with evol-instruct.
\newblock In \emph{Proceedings of ICLR}, 2024{\natexlab{c}}.
\newblock URL \url{https://openreview.net/forum?id=UnUwSIgK5W}.

\bibitem[{Meta LlaMa team}(2024)]{llama3modelcard}
{Meta LlaMa team}.
\newblock Llama 3 model card.
\newblock \url{https://github.com/meta-llama/llama3/blob/main/MODEL_CARD.md}, 2024.
\newblock Accessed: June 7, 2024.

\bibitem[OpenAI(2023{\natexlab{a}})]{abs-2303-08774}
OpenAI.
\newblock {GPT-4} technical report.
\newblock \emph{CoRR}, abs/2303.08774, 2023{\natexlab{a}}.
\newblock URL \url{https://doi.org/10.48550/arXiv.2303.08774}.

\bibitem[OpenAI(2023{\natexlab{b}})]{gpt35}
OpenAI.
\newblock Gpt-3.5 turbo fine-tuning and api updates.
\newblock \url{https://openai.com/index/gpt-3-5-turbo-fine-tuning-and-api-updates/}, 2023{\natexlab{b}}.
\newblock Accessed: June 7, 2024.

\bibitem[OpenAI(2024)]{gpt4o}
OpenAI.
\newblock Hello gpt-4o.
\newblock \url{https://openai.com/index/hello-gpt-4o/}, 2024.
\newblock Accessed: June 7, 2024.

\bibitem[Rozi{\`{e}}re et~al.(2023)Rozi{\`{e}}re, Gehring, Gloeckle, Sootla, Gat, Tan, Adi, Liu, Remez, Rapin, Kozhevnikov, et~al.]{abs-2308-12950}
Baptiste Rozi{\`{e}}re, Jonas Gehring, Fabian Gloeckle, Sten Sootla, Itai Gat, Xiaoqing~Ellen Tan, Yossi Adi, Jingyu Liu, Tal Remez, J{\'{e}}r{\'{e}}my Rapin, Artyom Kozhevnikov, et~al.
\newblock Code llama: Open foundation models for code.
\newblock \emph{CoRR}, abs/2308.12950, 2023.
\newblock URL \url{https://doi.org/10.48550/arXiv.2308.12950}.

\bibitem[Shao et~al.(2024)Shao, Li, Fei, Yan, Lin, and Qiu]{abs-2402-14526}
Yunfan Shao, Linyang Li, Zhaoye Fei, Hang Yan, Dahua Lin, and Xipeng Qiu.
\newblock Balanced data sampling for language model training with clustering.
\newblock In Lun{-}Wei Ku, Andre Martins, and Vivek Srikumar (eds.), \emph{Findings of ACL}, pp.\  14012--14023, 2024.
\newblock URL \url{https://doi.org/10.18653/v1/2024.findings-acl.833}.

\bibitem[Singh \& Strouse(2024)Singh and Strouse]{abs-2402-14903}
Aaditya~K. Singh and DJ~Strouse.
\newblock Tokenization counts: the impact of tokenization on arithmetic in frontier llms.
\newblock \emph{CoRR}, abs/2402.14903, 2024.
\newblock URL \url{https://doi.org/10.48550/arXiv.2402.14903}.

\bibitem[Sun et~al.(2024)Sun, Chen, Xu, Cheng, Ma, Yin, Wang, Han, Zhu, Yuan, et~al.]{abs-2403-14734}
Qiushi Sun, Zhirui Chen, Fangzhi Xu, Kanzhi Cheng, Chang Ma, Zhangyue Yin, Jianing Wang, Chengcheng Han, Renyu Zhu, Shuai Yuan, et~al.
\newblock A survey of neural code intelligence: Paradigms, advances and beyond.
\newblock \emph{CoRR}, abs/2403.14734, 2024.
\newblock URL \url{https://doi.org/10.48550/arXiv.2403.14734}.

\bibitem[Tian et~al.(2024)Tian, Ye, Qin, Cong, Lin, Pan, Wu, Hui, Liu, Liu, and Sun]{abs-2401-04621}
Runchu Tian, Yining Ye, Yujia Qin, Xin Cong, Yankai Lin, Yinxu Pan, Yesai Wu, Haotian Hui, Weichuan Liu, Zhiyuan Liu, and Maosong Sun.
\newblock Debugbench: Evaluating debugging capability of large language models.
\newblock In \emph{Findings of ACL}, pp.\  4173--4198, 2024.
\newblock URL \url{https://doi.org/10.18653/v1/2024.findings-acl.247}.

\bibitem[Touvron et~al.(2023)Touvron, Martin, Stone, Albert, Almahairi, Babaei, Bashlykov, Batra, Bhargava, Bhosale, et~al.]{abs-2307-09288}
Hugo Touvron, Louis Martin, Kevin Stone, Peter Albert, Amjad Almahairi, Yasmine Babaei, Nikolay Bashlykov, Soumya Batra, Prajjwal Bhargava, Shruti Bhosale, et~al.
\newblock Llama 2: Open foundation and fine-tuned chat models.
\newblock \emph{CoRR}, abs/2307.09288, 2023.
\newblock URL \url{https://doi.org/10.48550/arXiv.2307.09288}.

\bibitem[Vadlamani et~al.(2021)Vadlamani, Kalicheti, and Chimalakonda]{VadlamaniKC21}
Aparna Vadlamani, Rishitha Kalicheti, and Sridhar Chimalakonda.
\newblock Apiscanner - towards automated detection of deprecated apis in python libraries.
\newblock In \emph{Proceedings of ICSE}, pp.\  5--8, 2021.
\newblock URL \url{https://doi.org/10.1109/ICSE-Companion52605.2021.00022}.

\bibitem[Wang et~al.(2020)Wang, Li, Liu, and Cai]{Wang0LC20}
Jiawei Wang, Li~Li, Kui Liu, and Haipeng Cai.
\newblock Exploring how deprecated python library apis are (not) handled.
\newblock In \emph{Proceedings of ESEC/FSE}, pp.\  233--244. {ACM}, 2020.
\newblock URL \url{https://doi.org/10.1145/3368089.3409735}.

\bibitem[Wu et~al.(2022)Wu, Caccia, Li, Li, Qi, and Haffari]{WuCLLQH22}
Tongtong Wu, Massimo Caccia, Zhuang Li, Yuan{-}Fang Li, Guilin Qi, and Gholamreza Haffari.
\newblock Pretrained language model in continual learning: {A} comparative study.
\newblock In \emph{ICLR}. OpenReview.net, 2022.
\newblock URL \url{https://openreview.net/forum?id=figzpGMrdD}.

\bibitem[Wu et~al.(2024)Wu, Luo, Li, Pan, Vu, and Haffari]{abs-2402-01364}
Tongtong Wu, Linhao Luo, Yuan{-}Fang Li, Shirui Pan, Thuy{-}Trang Vu, and Gholamreza Haffari.
\newblock Continual learning for large language models: {A} survey.
\newblock \emph{CoRR}, abs/2402.01364, 2024.
\newblock URL \url{https://doi.org/10.48550/arXiv.2402.01364}.

\bibitem[Yadav et~al.(2023)Yadav, Sun, Ding, Li, Zhang, Tan, Bhatia, Ma, Nallapati, Ramanathan, et~al.]{yadav-etal-2023-exploring}
Prateek Yadav, Qing Sun, Hantian Ding, Xiaopeng Li, Dejiao Zhang, Ming Tan, Parminder Bhatia, Xiaofei Ma, Ramesh Nallapati, Murali~Krishna Ramanathan, et~al.
\newblock Exploring continual learning for code generation models.
\newblock In \emph{Proceedings of ACL}, pp.\  782--792, 2023.
\newblock URL \url{https://aclanthology.org/2023.acl-short.68}.

\bibitem[Yan et~al.(2023)Yan, Tian, Li, Chen, and Wang]{YanTLCW23}
Weixiang Yan, Yuchen Tian, Yunzhe Li, Qian Chen, and Wen Wang.
\newblock Codetransocean: {A} comprehensive multilingual benchmark for code translation.
\newblock In \emph{Findings of EMNLP}, pp.\  5067--5089, 2023.
\newblock URL \url{https://doi.org/10.18653/v1/2023.findings-emnlp.337}.

\bibitem[Yao et~al.(2018)Yao, Weld, Chen, and Sun]{YaoWCS18}
Ziyu Yao, Daniel~S. Weld, Wei{-}Peng Chen, and Huan Sun.
\newblock Staqc: {A} systematically mined question-code dataset from stack overflow.
\newblock In \emph{Proceedings of WebConf}, pp.\  1693--1703, 2018.
\newblock URL \url{https://doi.org/10.1145/3178876.3186081}.

\bibitem[Yin et~al.(2018)Yin, Deng, Chen, Vasilescu, and Neubig]{YinDCVN08}
Pengcheng Yin, Bowen Deng, Edgar Chen, Bogdan Vasilescu, and Graham Neubig.
\newblock Learning to mine aligned code and natural language pairs from stack overflow.
\newblock In \emph{Proceedings of ICMSR}, pp.\  476--486, 2018.
\newblock URL \url{https://doi.org/10.1145/3196398.3196408}.

\bibitem[Yu et~al.(2024)Yu, Shen, Ran, Zhang, Zhang, Ma, Liang, Li, Wang, and Xie]{YuSRZZMLLWX24}
Hao Yu, Bo~Shen, Dezhi Ran, Jiaxin Zhang, Qi~Zhang, Yuchi Ma, Guangtai Liang, Ying Li, Qianxiang Wang, and Tao Xie.
\newblock Codereval: {A} benchmark of pragmatic code generation with generative pre-trained models.
\newblock In \emph{Proceedings of ICSE}, pp.\  37:1--37:12, 2024.
\newblock URL \url{https://doi.org/10.1145/3597503.3623316}.

\bibitem[Zhang et~al.(2023)Zhang, Zhang, Zhai, Fang, Yu, Sun, and Chen]{abs-2310-08879}
Quanjun Zhang, Tongke Zhang, Juan Zhai, Chunrong Fang, Bowen Yu, Weisong Sun, and Zhenyu Chen.
\newblock A critical review of large language model on software engineering: An example from chatgpt and automated program repair.
\newblock \emph{CoRR}, abs/2310.08879, 2023.
\newblock URL \url{https://doi.org/10.48550/arXiv.2310.08879}.

\bibitem[Zhang et~al.(2021)Zhang, Yang, Xia, Lo, Ren, and Grundy]{ZhangY00RG21}
Zejun Zhang, Yanming Yang, Xin Xia, David Lo, Xiaoxue Ren, and John~C. Grundy.
\newblock Unveiling the mystery of {API} evolution in deep learning frameworks: {A} case study of tensorflow 2.
\newblock In \emph{Proceedings of ICSE}, pp.\  238--247, 2021.
\newblock URL \url{https://doi.org/10.1109/ICSE-SEIP52600.2021.00033}.

\bibitem[Zhang et~al.(2020)Zhang, Zhu, Wen, Tao, Liu, and Xiong]{ZhangZWTLX20}
Zhaoxu Zhang, Hengcheng Zhu, Ming Wen, Yida Tao, Yepang Liu, and Yingfei Xiong.
\newblock How do python framework apis evolve? an exploratory study.
\newblock In \emph{Proceedings of SANER}, pp.\  81--92, 2020.
\newblock URL \url{https://doi.org/10.1109/SANER48275.2020.9054800}.

\bibitem[Zhao et~al.(2024)Zhao, Brumbaugh, Wang, Hajishirzi, and Smith]{abs-2402-16797}
Bowen Zhao, Zander Brumbaugh, Yizhong Wang, Hannaneh Hajishirzi, and Noah~A. Smith.
\newblock Set the clock: Temporal alignment of pretrained language models.
\newblock In \emph{Findings of ACL}, pp.\  15015--15040, 2024.
\newblock URL \url{https://doi.org/10.18653/v1/2024.findings-acl.892}.

\bibitem[Zheng et~al.(2023)Zheng, Xia, Zou, Dong, Wang, Xue, Wang, Shen, Wang, Li, et~al.]{abs-2303-17568}
Qinkai Zheng, Xiao Xia, Xu~Zou, Yuxiao Dong, Shan Wang, Yufei Xue, Zihan Wang, Lei Shen, Andi Wang, Yang Li, et~al.
\newblock Codegeex: {A} pre-trained model for code generation with multilingual evaluations on humaneval-x.
\newblock \emph{CoRR}, abs/2303.17568, 2023.
\newblock URL \url{https://doi.org/10.48550/arXiv.2303.17568}.

\bibitem[Zhu et~al.(2022{\natexlab{a}})Zhu, Jain, Suresh, Ravindran, Tipirneni, and Reddy]{abs-2206-08474}
Ming Zhu, Aneesh Jain, Karthik Suresh, Roshan Ravindran, Sindhu Tipirneni, and Chandan~K. Reddy.
\newblock Xlcost: {A} benchmark dataset for cross-lingual code intelligence.
\newblock \emph{CoRR}, abs/2206.08474, 2022{\natexlab{a}}.
\newblock \doi{10.48550/ARXIV.2206.08474}.
\newblock URL \url{https://doi.org/10.48550/arXiv.2206.08474}.

\bibitem[Zhu et~al.(2022{\natexlab{b}})Zhu, Suresh, and Reddy]{Zhu0R22}
Ming Zhu, Karthik Suresh, and Chandan~K. Reddy.
\newblock Multilingual code snippets training for program translation.
\newblock In \emph{Proceedings of AAAI}, pp.\  11783--11790, 2022{\natexlab{b}}.
\newblock URL \url{https://doi.org/10.1609/aaai.v36i10.21434}.

\bibitem[Zhuo et~al.(2024)Zhuo, Vu, Chim, Hu, Yu, Widyasari, Yusuf, Zhan, et~al.]{abs-2406-15877}
Terry~Yue Zhuo, Minh~Chien Vu, Jenny Chim, Han Hu, Wenhao Yu, Ratnadira Widyasari, Imam Nur~Bani Yusuf, Haolan Zhan, et~al.
\newblock Bigcodebench: Benchmarking code generation with diverse function calls and complex instructions.
\newblock \emph{CoRR}, abs/2406.15877, 2024.
\newblock URL \url{https://doi.org/10.48550/arXiv.2406.15877}.

\end{thebibliography}
